\documentclass[a4paper,10pt,twoside]{cpc-hepnp}
\usepackage{CJK,upgreek,fancyhdr}
\usepackage{multicol}
\usepackage{graphicx}
\usepackage{booktabs}
\usepackage{amssymb,bm,mathrsfs,bbm,amscd}
\usepackage[tbtags]{amsmath}
\usepackage{lastpage}

\begin{document}
\begin{CJK*}{GB}{gbsn}

\fancyhead[c]{\small Chinese Physics C~~~Vol. xx, No. x (201x) xxxxxx}
\fancyfoot[C]{\small 010201-\thepage}

\footnotetext[0]{Received 31 June 2015}

\title{Upgrade of Beam Energy Measurement System at BEPC-II\thanks{This work
is supported in part by National Natural Science Foundation of China (NSFC)
under contracts Nos.: 11375206, 10775142, 10825524, 11125525, 11235011; the
Ministry of Science and Technology of China under Contract Nos.: 2015CB856700,
2015CB856706; State key laboratory of particle and detection and electronics;
and the CAS Center for Excellence in Particle Physics (CCEPP),
the RFBR grant No 14-02-00129-a. Part of this work related to the design of
ZnSe viewports is supported by the Russian Science Foundation (project No
14-50-00080)}}

\author{%
      Jian-Yong Zhang (ÕŽ¨ÓÂ)$^{1;1)}$\email{jyzhang@mail.ihep.ac.cn}
\quad Xiao Cai (²ÌÐ¥)$^{1}$
\quad Xiao-Hu Mo (ĪÏþ»¢)$^{1;2)}$\email{moxh@mail.ihep.ac.cn}
\quad Di-Zhou Guo (¹ùµÏÖÛ)$^{1}$  \\
\quad Jian-Li Wang (Íõ½¨Á¦)$^{1}$
\quad Bai-Qi Liu (Áõ°ÛÆæ)$^{1}$
\quad M. N. Achasov$^{2,3}$
\quad A. A. Krasnov$^{2,3}$
\quad N.Yu. Muchnoi$^{2,3}$ \\
\quad E. E. Pyata$^{2}$
\quad E. V. Mamoshkina$^{2}$
\quad F. A. Harris$^{4}$
}
\maketitle

\address{%
$^1$ Institute of High Energy Physics, Chinese Academy of Sciences, Beijing 100049, China\\
$^2$ Budker Institute of Nuclear Physics, Siberian Branch of the Russian Academy of Sciences,
11 Lavrentyev, Novosibirsk 630090, Russia\\
$^3$ Novosibirsk State University, Novosibirsk 630090, Russia\\
$^4$ University of Hawaii, Honolulu, Hawaii 96822, USA
}

\begin{abstract}
The beam energy measurement system is of great importance and profit for both BEPC-II accelerator and  BES-III detector. The system is based on measuring the energies of Compton back-scattered photons. Many advanced techniques and precise instruments are employed to realize the highly accurate measurement of positron/electron beam energy. During five year's running period, in order to meet the requirement of data taking and improve the capacity of measurement itself, the upgradation of system is continued, which involve the component reformation of laser and optics subsystem,
replacement of view-port of the laser to vacuum insertion subsystem, the purchase of electric cooling system for high purity germanium detector, and the improvement of data acquisition and processing subsystem. The upgrading of system guarantees the smooth and efficient measuring of beam energy at BEPC-II and accommodates the accurate offline energy values for further physics analysis at BES-III.
\end{abstract}



\footnotetext[0]{\hspace*{-3mm}\raisebox{0.3ex}{$\scriptstyle\copyright$}2013
Chinese Physical Society and the Institute of High Energy Physics
of the Chinese Academy of Sciences and the Institute
of Modern Physics of the Chinese Academy of Sciences and IOP Publishing Ltd}%

\begin{multicols}{2}

\section{Introduction}

The upgraded Beijing electron-positron collider (BEPC-II) is a $\tau$-charm factory with a center mass of energy ranging from 2.0 to 4.6 GeV and a design peak luminosity of
$10^{33}$~cm$^{-2}$ s$^{-1}$~\cite{bepc,bii} at the center of mass energy 3.770 GeV. The upgraded Beijing spectrometer detector (BES-III) with high efficiency and good resolution for both charged and neutral particles was constructed and started data taking in 2008~\cite{bes,yellow}.

After colossal amounts of data are acquired and analyzed, the statistical uncertainties in physics analysis become smaller and smaller, while the systematic uncertainties play more and more prominent roles~\cite{sim1,sim2,sim3}, one of which is the uncertainty due to the measurement of beam energy.
To decrease such an uncertainty, start from year 2007, a high accuracy beam energy measurement system (BEMS) located at the north crossing point (NCP) of BEPC-II was designed, constructed, and put into the commissioning at the end of 2010~\cite{bems2009,bems2010,bems,bems2}. The launching of system is excellently well, two days are utilized to perform $\psi'$ scan. The mass difference between the PDG2010 value and the measured result by BEMS is $1 \pm 36$ keV, the deviation of which indicates that the relative accuracy of BEMS is at the level of 2 $\times 10^{-5}$~\cite{bems}.

The establishment of BEMS improves the measurement capacity of both accelerator and detector, and can provide the beam energy, energy spread and their corresponding errors, all of which are crucial and useful information for the physical analysis of BES-III and the luminosity tuning at  BEPC-II. The first exhibition of such a capacity of BEMS is the test scan of $\tau$ mass that was performed at the end of 2011. The integrated luminosity of $\tau$ sample is 23.26 pb$^{-1}$, the mass of $\tau$ lepton is determined as $m_{\tau} = 1776.91 \pm 0.12 ^{ + 0.10} _{- 0.13}$ MeV~\cite{taupaper}, among which the systematic uncertainty due to energy scale is less than 0.09 MeV.

During five years running, BEMS participated the various data collections at BES-III, that include $J/\psi$ and $\psi'$ resonance samples, R value scan sample, the high excited charmonium states samples and so forth. The high precision beam energy values are measured and accommodated for the offline data analysis.

The high precision of energy calibration acquired by BEMS is actually based on Compton backscattering principle. The working scheme of this system can be recapitulated as follows~\cite{principle}: firstly, the laser source provides the laser beam and the optics system focuses the laser beam and guides it to make head-on collisions with the electron (or positron) beam in the vacuum pipe, here the Compton backscattering process happens; after that the backscattering high energy photon will be detected by the HPGe detector. The more engineering details can be found in Ref.~\cite{bems}. As a matter of fact, many advanced techniques and precise instruments are employed to achieve such a highly accurate measurement of beam energy. The whole system can be sub-divided into four parts according to their technique and engineering characters: 1) Laser source and optics system which supply low energy laser and focused photons; 2) Laser to vacuum insertion system where laser beam collides with electron or positron beam; 3) High Purity Germanium (HPGe) detector to measure  backscattering high energy photons; 4) Data acquisition and running control system for information processing and analyzing. The layout schematic of the system is shown in Fig.~\ref{ln2fill}.

\begin{center}
\begin{minipage}{8cm}
\includegraphics[width=8cm]{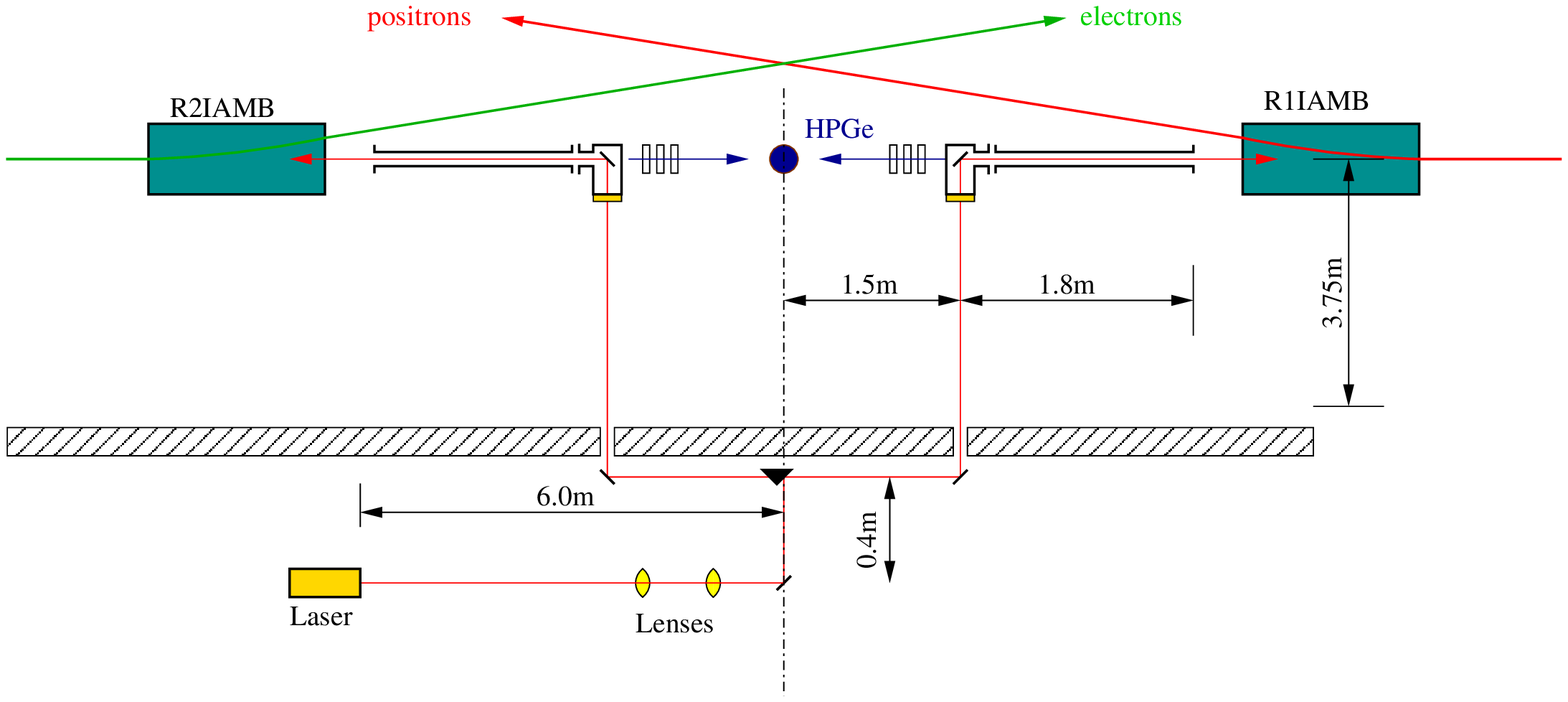}
\end{minipage}
\figcaption{\label{ln2fill}
Simplified schematic of beam energy measurement system. The positron and electron beams are indicated. R1IAMB and R2IAMB are accelerator magnets, and the HPGe detector is represented by the dot at center. The shielding wall of the beam tunnel is shown cross-hatched, and the laser is located out side the tunnel.
}
\end{center}

However, because of the physics requirement or the request of more precise measurement, the updates for BEMS have been performing since its commissioning. The details are depicted in the following sections. The laser and optical system upgrade is introduced in section 2.
The improvement of laser to vacuum insertion system is presented in section 3. 
The update of detection system is introduced in section 4. 
The data acquisition and process system is presented subsequently. The interlock system is introduced in the last section but one, after it there is a short summary.

\section{Laser and optical system}
This system consists of laser source supplying low energy photons, two lenses focusing the beam light, and three reflecting mirrors and one movable prism directing the laser beam into the storage ring tunnel. All relevant instruments are installed along the shielding concrete wall as shown in Fig.~\ref{ln2fill}. The upgrading processes are described chronologically below.
\begin{center}
\begin{minipage}{4.cm}
\includegraphics[height=4.cm]{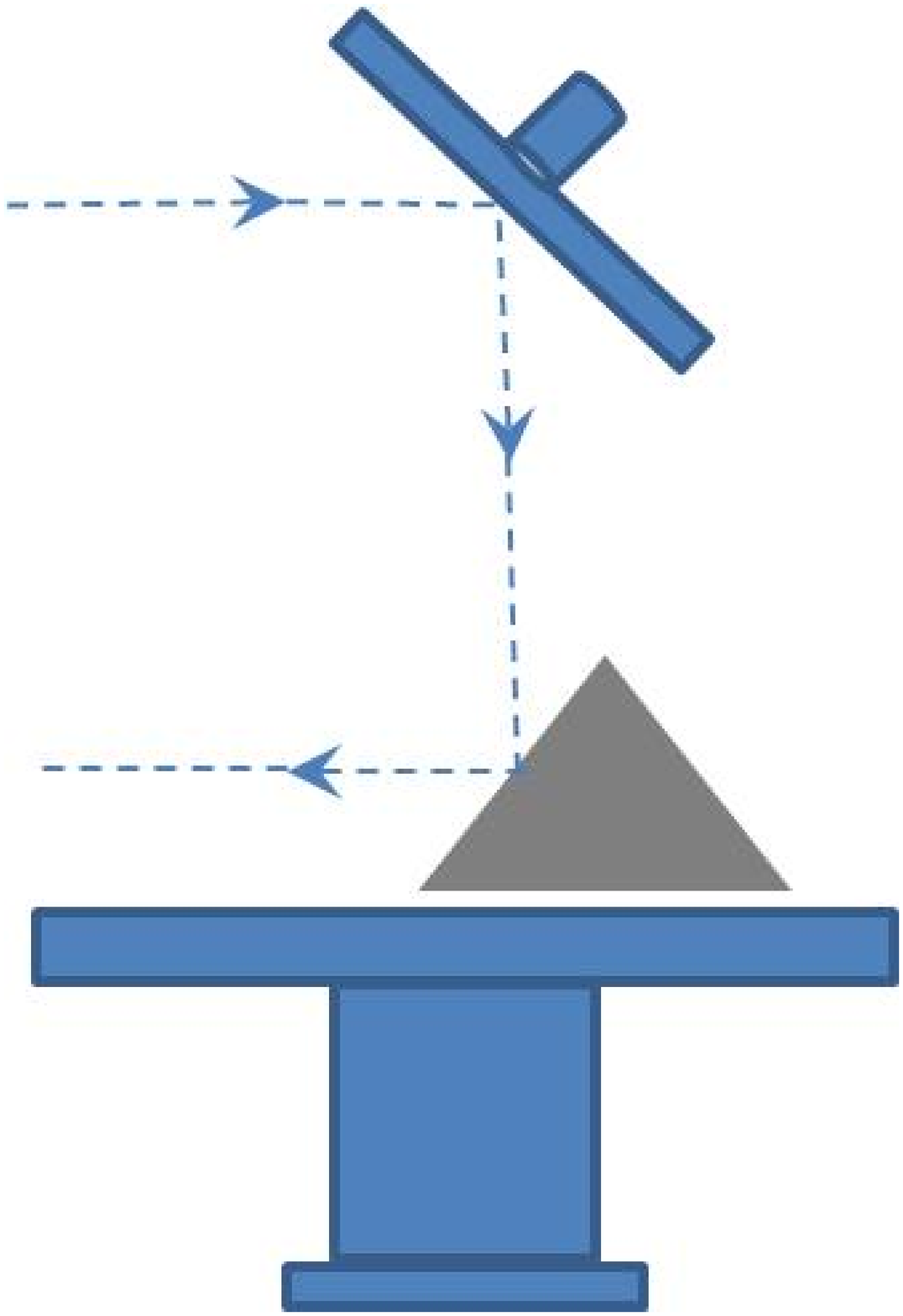}
\centerline{(a)
}
\end{minipage}
\begin{minipage}{4.cm}
\includegraphics[height=4.cm]{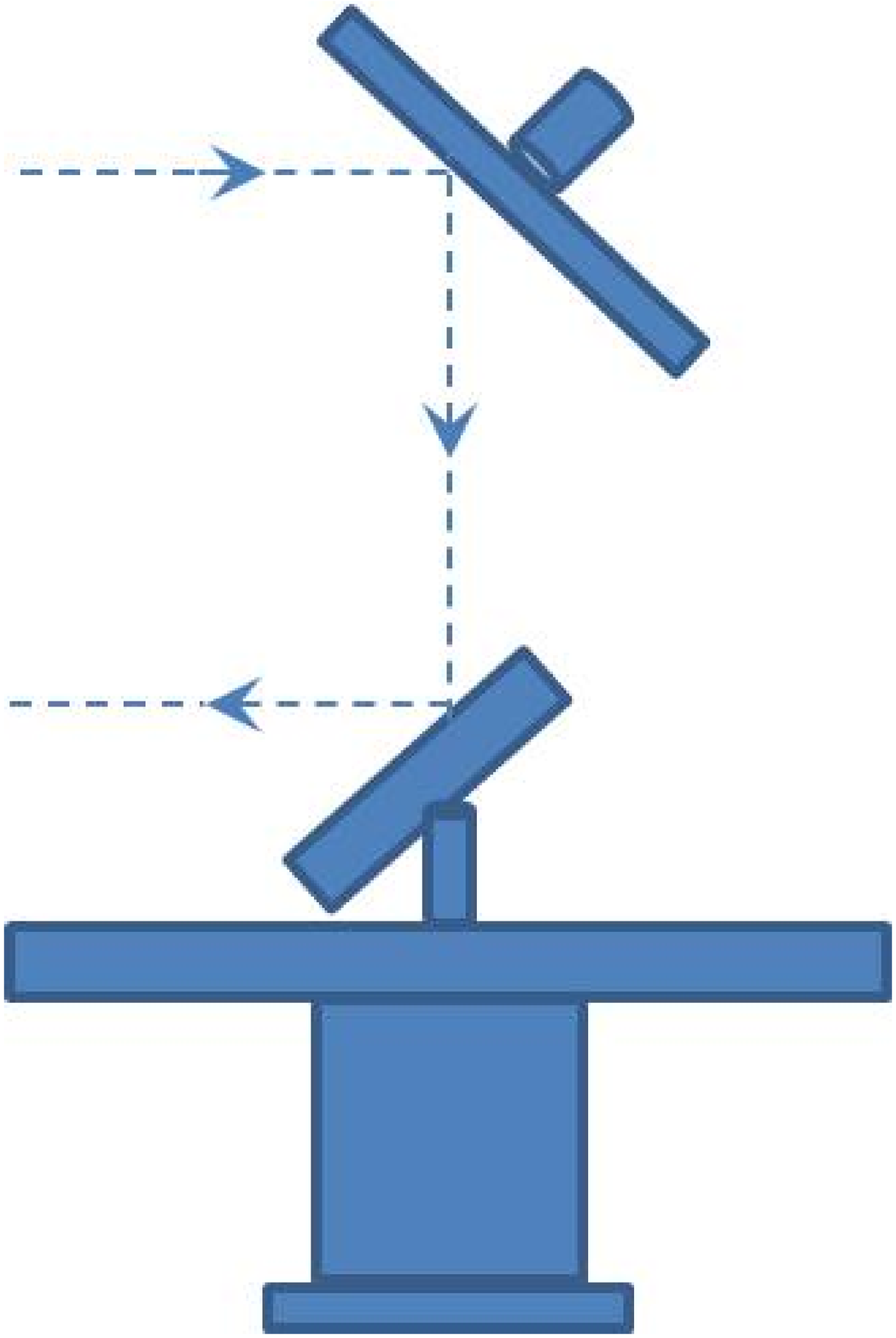}
\centerline{(d)
}
\end{minipage}
\begin{minipage}{4.cm}
\includegraphics[height=4.cm]{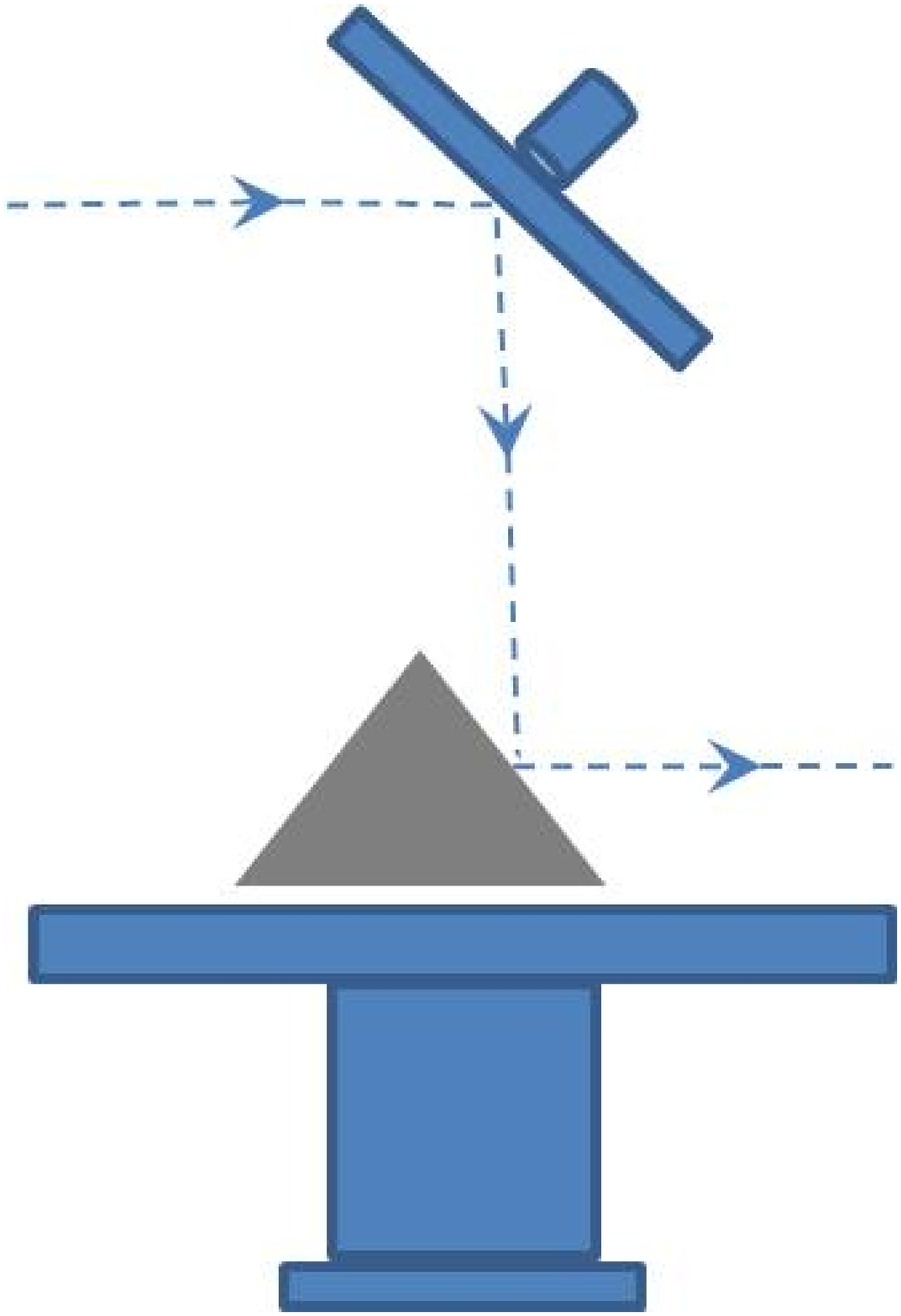}
\centerline{(b)
}
\end{minipage}
\begin{minipage}{4.cm}
\includegraphics[height=4.cm]{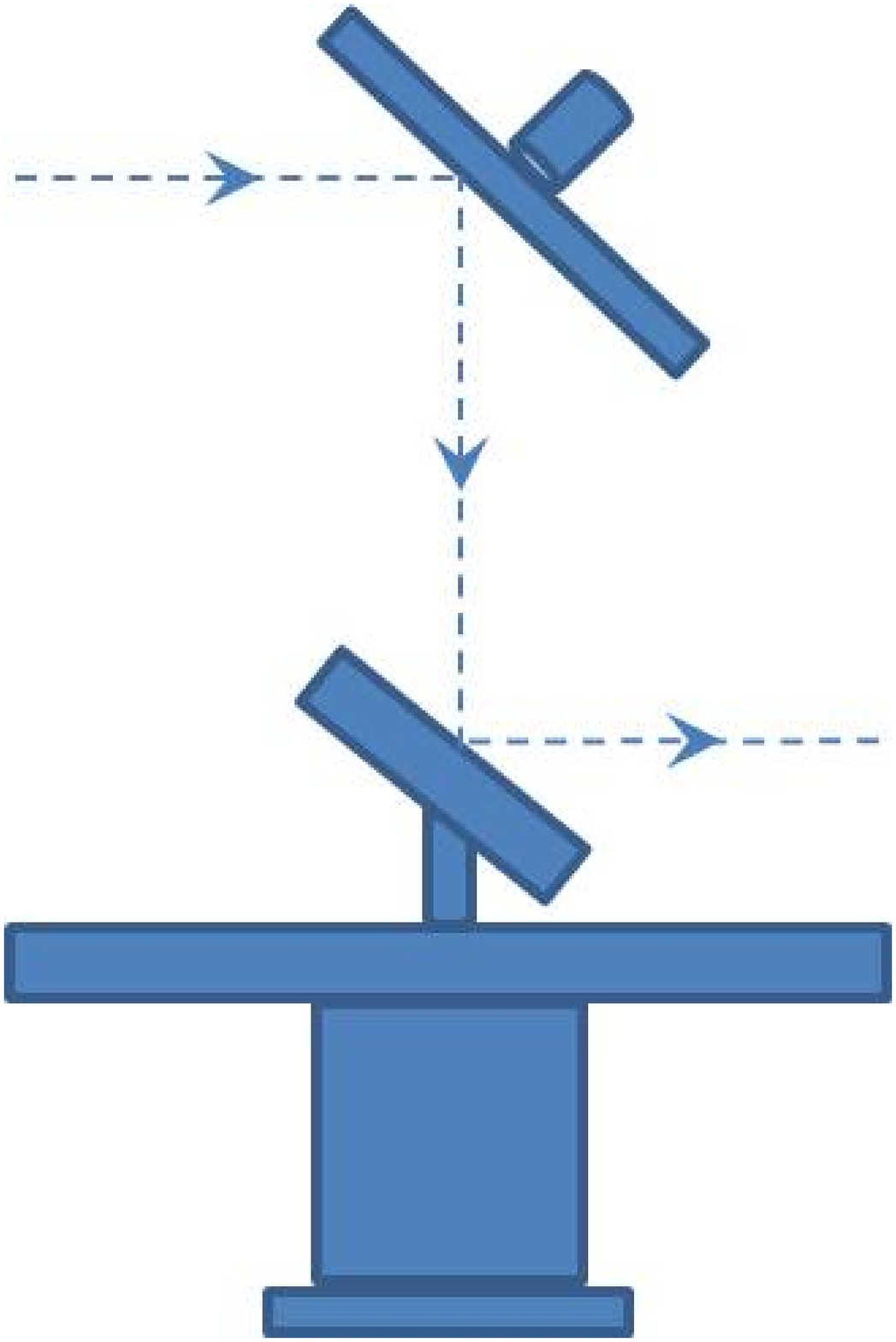}
\centerline{(e)
}
\end{minipage}
\begin{minipage}{4cm}
\includegraphics[width=3.5cm]{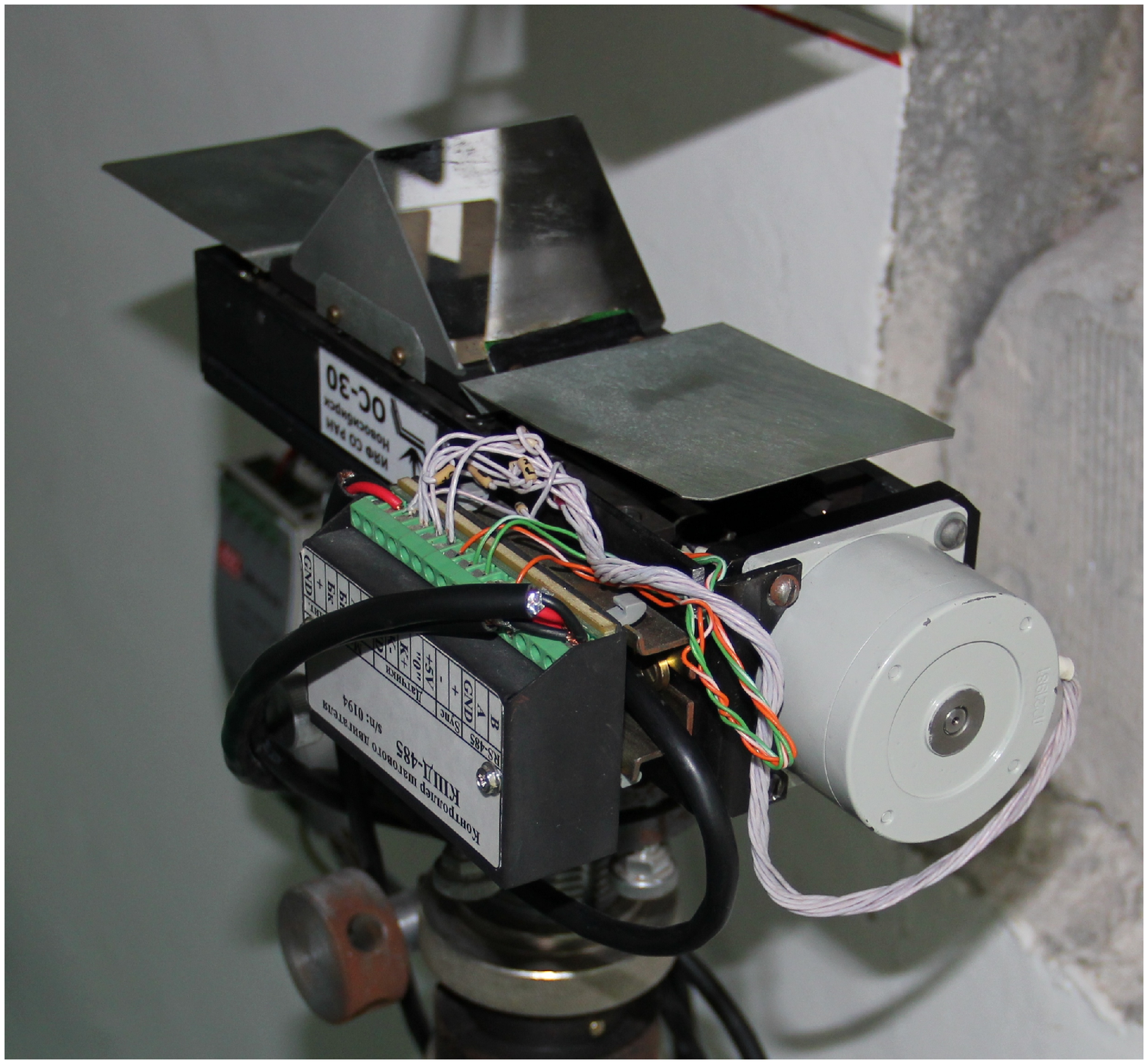}
\centerline{(c)
}
\end{minipage}
\begin{minipage}{4cm}
\includegraphics[width=4.0cm]{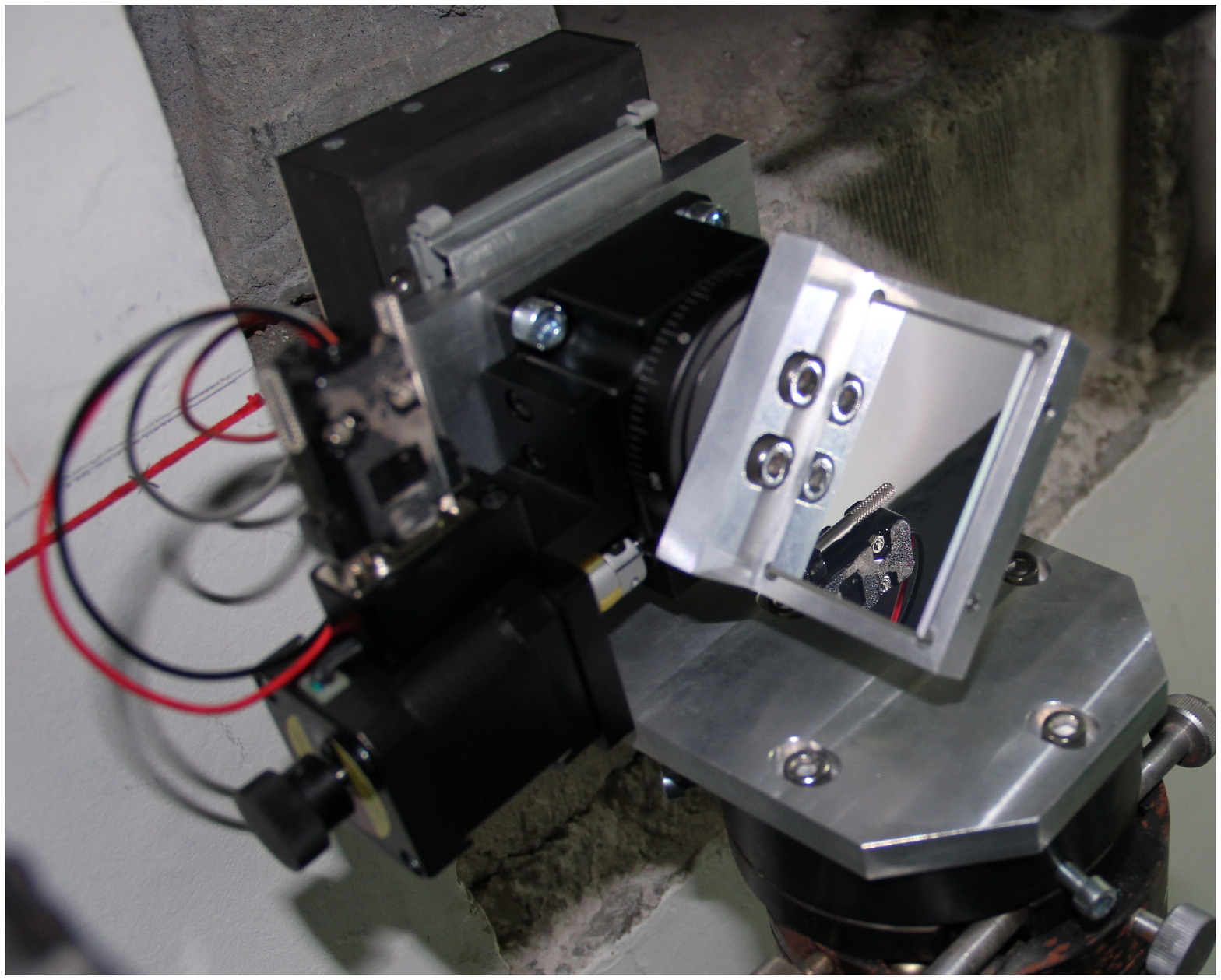}
\centerline{(f)}
\end{minipage}
\end{center}
\figcaption{\label{mov}The left side is the translatory prism and the right case is the rotary platform. The laser light, represented by the dash line, is reflected to the electron (a, d) and positron (b, e) side using translatory prism and rotary platform respectively. The assembly pictures of
translatory prism and rotary platform are shown in (c) and (f) separately.}

Firstly, it is the renewal of lenses (denoted by ``lenses'' in Fig.~\ref{ln2fill}).
The laser beam is focused at the BEPC-II vacuum chamber entrance flange, where the geometrical aperture is minimal: vertical size $\times$ horizontal size is 14 mm $\times$ 50 mm. Since the total distance from the laser output aperture to the entrance flange of the BEPC-II vacuum chamber is about 18 m, the focusing function can hardly be realized by a single lenses. Therefore a doublet of lenses with focal lengths of 40 cm are placed respectively at distance 300.0 and 381.6 cm from the laser  output window and provide the laser beam transverse size at the flange from 2.0 to 2.5 mm.

The original lenses is made of zinc selenide (ZnSe), with the laser transmissivity of 69\% and 61\% for two lenses respectively. The synthetic transmissivity is merely 42\%, which means more than half of photons are consumed during the focusing. Two new ZnSe lenses are manufactured by BINP group, whose focus length is 42 cm with the transmissivity of 99\% each. The synthetic transmissivity is up to 98\% and the corresponding positions are 330 cm and 415 cm before the laser output window separately.

Secondly, it is the replacement of prism.
During the operation of BEMS, the energy of positron beam or electron beam is measured alternately. The alternation is executed by a translatory reflector prism, as shown in Fig.~\ref{mov}, where (a) and (b) are the schematic explanation for beam path selection, and (c) is the physical picture of prism. The laser beam is directed by the prism towards the right or left mirror to collide with position or electron beam automatically. During the service, the dust accumulated in the groove of prism support gives a rise to the braking effect on the automatical movement and eventually leads to the immobilization of prism. The switch between electron and positron has to be manipulated manually, which is actually impractical for the full-day continuous measurement.

Instead of changing the old prism with a new one, a brand new design is put forth to prevent recurrence of the similar incident. A rotary platform shown in Fig.~\ref{mov}(f) is investigated. The rotary gear is covered below the platform to ensure durable. A through hole in the center of the rotary platform satisfies the coaxial requirement strictly with the center of rotation. The center aperture of the rotary platform suffices the strict tolerance limitation, both of which will be benefit to the precise positioning. The absolute positioning accuracy of the rotary platform is 0.005 degree. A special support is designed, forged, and installed to fix the rotary platform. A reflect mirror with its frame is fixed on the rotary platform  to change the direction of the laser. The rotary platform is controlled by the step motor, whose step is 1.5$\times 10 ^{-6}$ rad, to rotate certain degrees precisely, so that the mirror will reflect the laser to the right or left side to make collision with positron or electron beam.

Lastly, it is about the issue of laser.
The source of initial photons is the GEM Selected 50$^{\mbox{\scriptsize TM}}$ $CO_2$ laser from Coherent, {\em Inc.} It is a continuous operation (CW), high power, and single-line narrow-width laser. It provides 25 W of CW power at the wavelength $\lambda_0=10.835231$ $\mu$m ( $\gamma$-quantum energy $\omega_0=0.114426901$ eV), which corresponds to 10P42 transition in the carbon dioxide molecule~\cite{CO2lambda}. The relative accuracy of $\omega_0$ at the level of better than 0.1~ppm is accessible.

After five years running, the grating coat becomes too thin to reflect photons, the laser was returned to the manufactory for repairing. Considering the designed beam current of BEPC-II is 910 mA, the requested physics energy region ($1.0 \sim 2.3$ GeV) is wide, the background near the NCP is complicated, a more powerful laser is indeed needed. Therefore, a more powerful laser with output power up to 50 W is chosen as the laser source. The power of the new laser is about twice of the old one, the Compton edge will be easier to obtain and the data taken time for measurement will be shorten. For the new laser, the wavelength $\lambda_0=10.591035$ $\mu$m ( $\gamma$-quantum energy $\omega_0=0.117065228$ eV), which corresponds to 10P20 transition in the carbon dioxide molecule \cite{CO2lambda}.

\begin{center}
\includegraphics[width=6.cm]{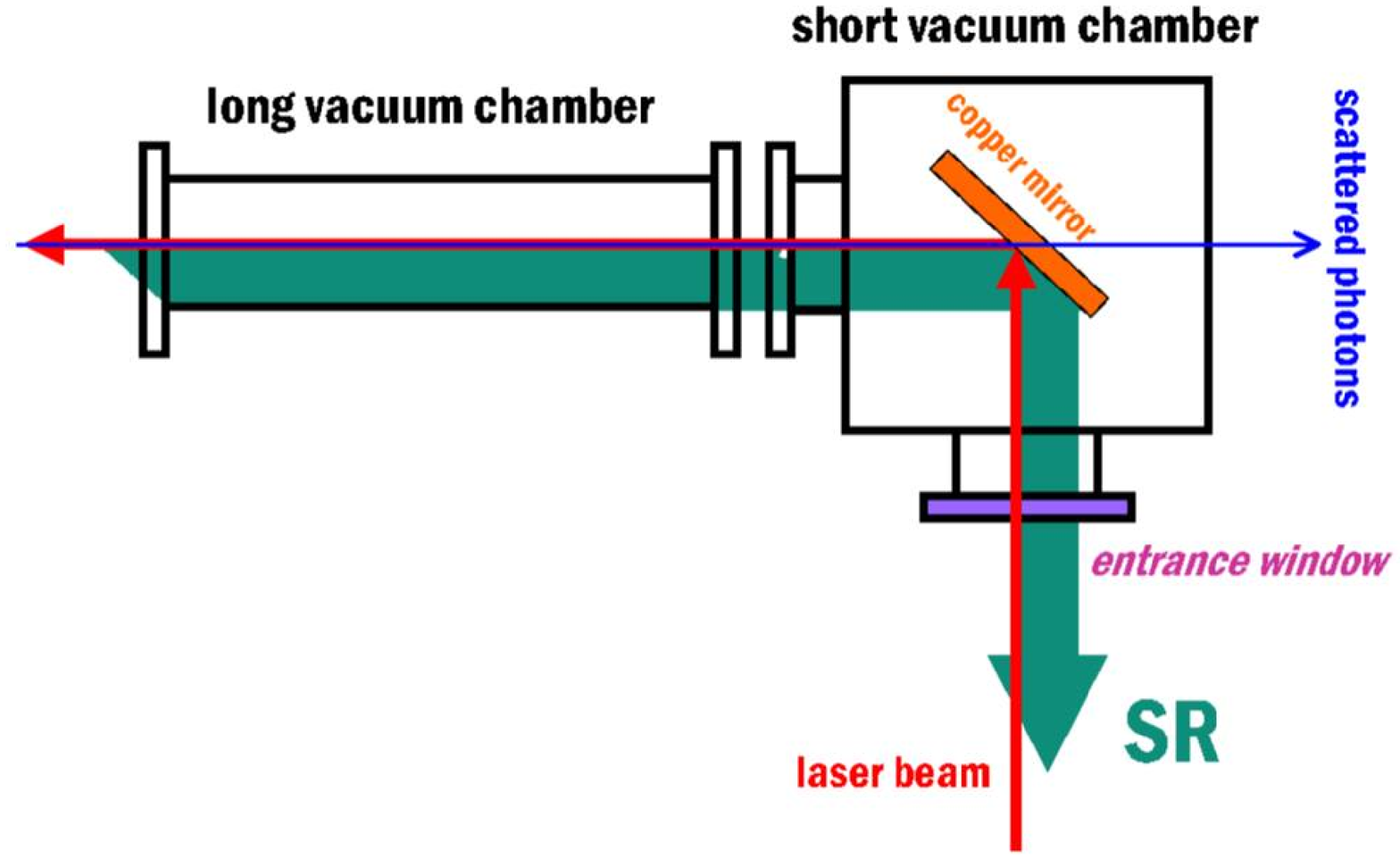}
\caption{\label{ltov}Schematic diagram of the laser to vacuum insertion of BEMS. SR represents the Synchrotron Radiation light.}
\end{center}

\section{Laser to vacuum insertion system}\label{sxt:insertion}

The laser to vacuum insertion system \cite{vacsys} is a crucial part of BEMS.
As shown in Fig.~\ref{ltov},
the system is composed of a special stainless vacuum chamber with entrance viewport and a reflect copper mirror. From this part, the laser beam is inserted into the vacuum chamber, reflected 90 degrees by the copper mirror and makes the head-on collision with the positron or electron beam.
Then the back-scattered photons pass through the copper mirror and are detected by the HPGe detector. Since the mirror is heated by reflecting laser and synchrotron light, it must be cooled by cooling water. In addition, since this part responds for connecting with the beam pipe of the storage ring, after installation, the vacuum chamber must be baked out at 250 degree for 24 hours in order to guarantee that the vacuum pressure must be better than 2$\times 10^{-10}$ Torr. As to this part, two advanced techniques are adopted and the first one is the synchrotron light alignment technique.

\begin{center}
\begin{minipage}{4cm}
\includegraphics[height=4cm,width=4cm]{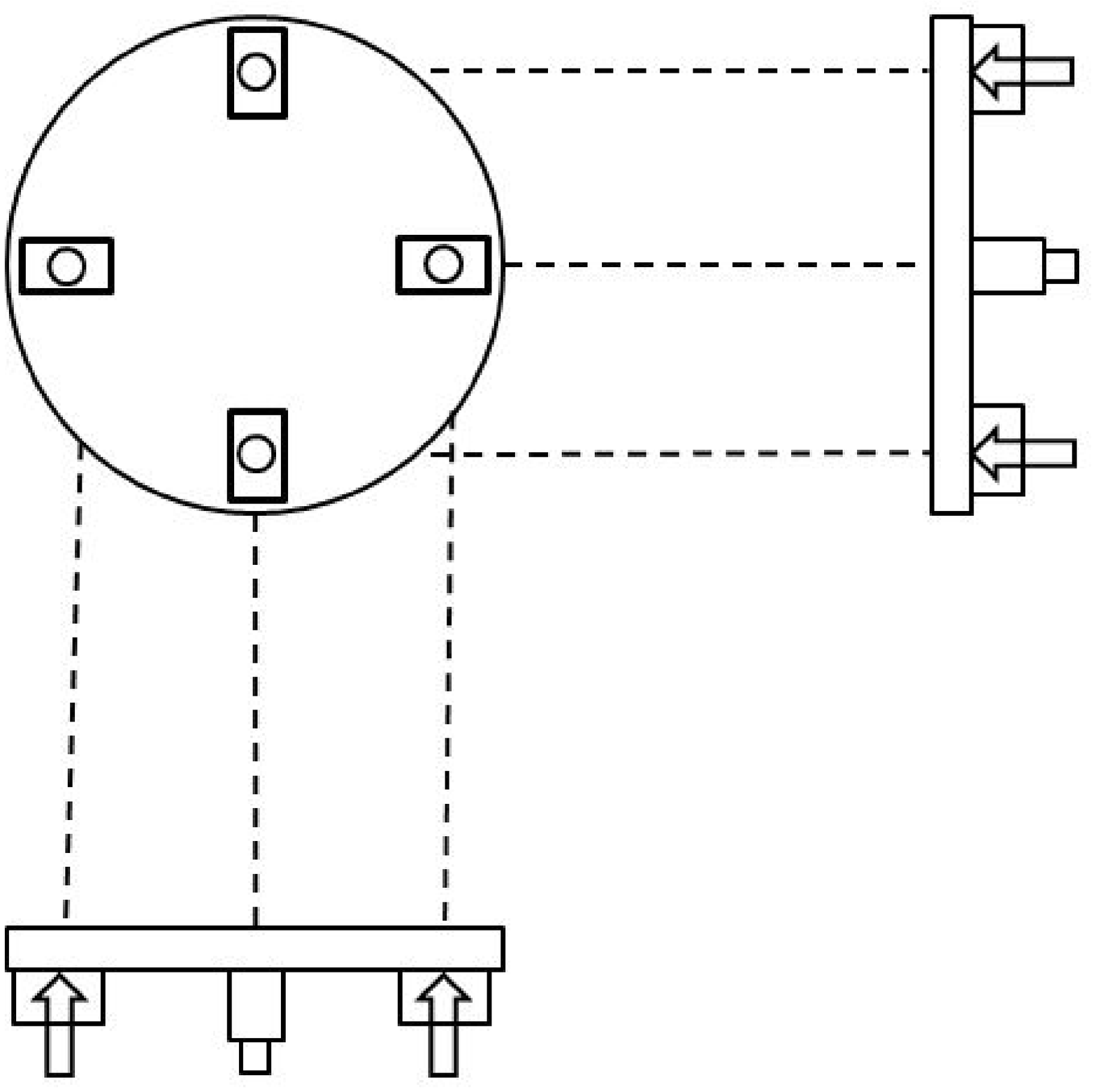}
\centerline{(a)}
\end{minipage}
\begin{minipage}{4cm}
\includegraphics[width=5cm]{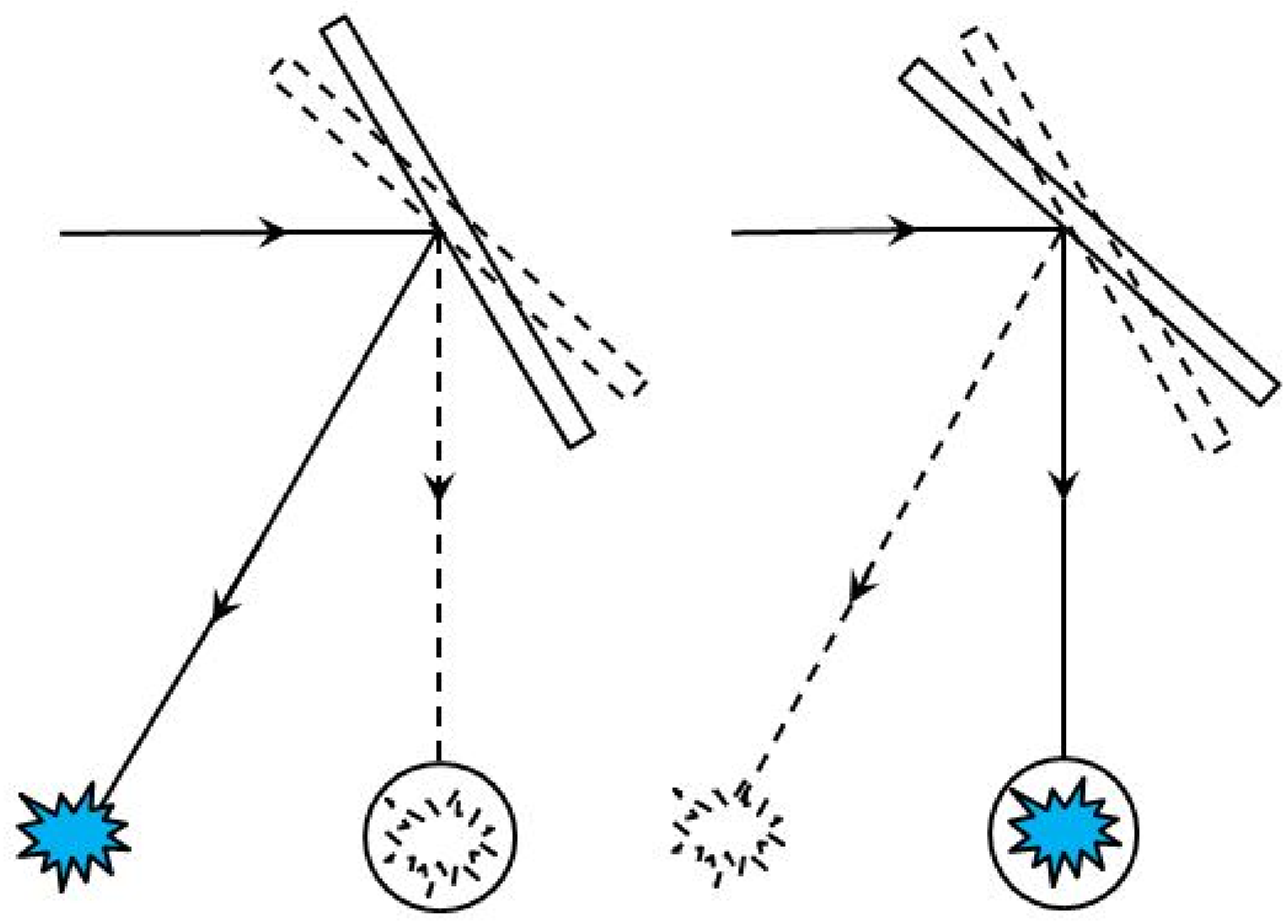}
\centerline{(b)}
\end{minipage}
\begin{minipage}{4cm}
\includegraphics[height=3.5cm]{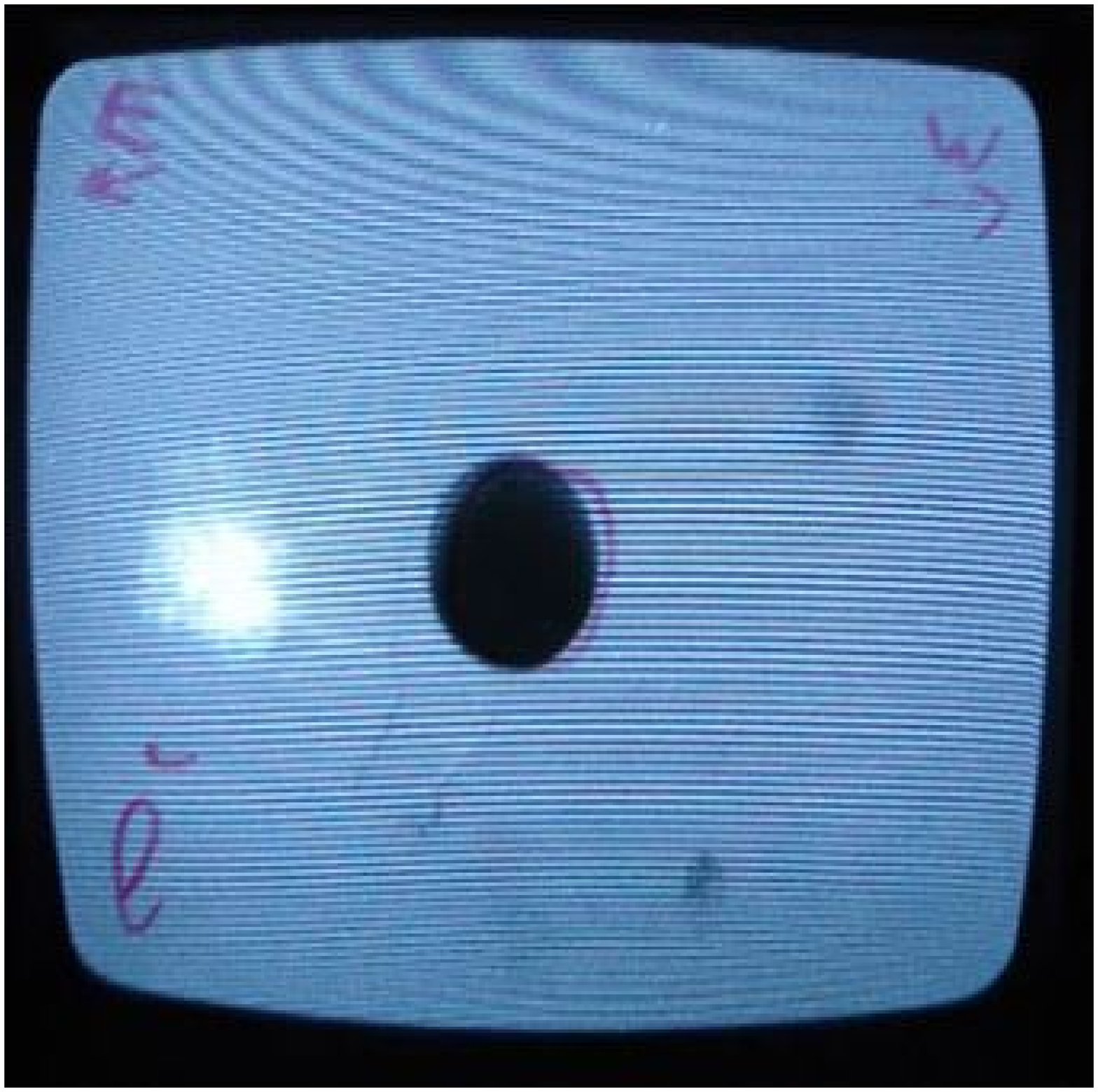}
\centerline{ (c)  }
\end{minipage}
\begin{minipage}{4cm}
\includegraphics[height=3.5cm]{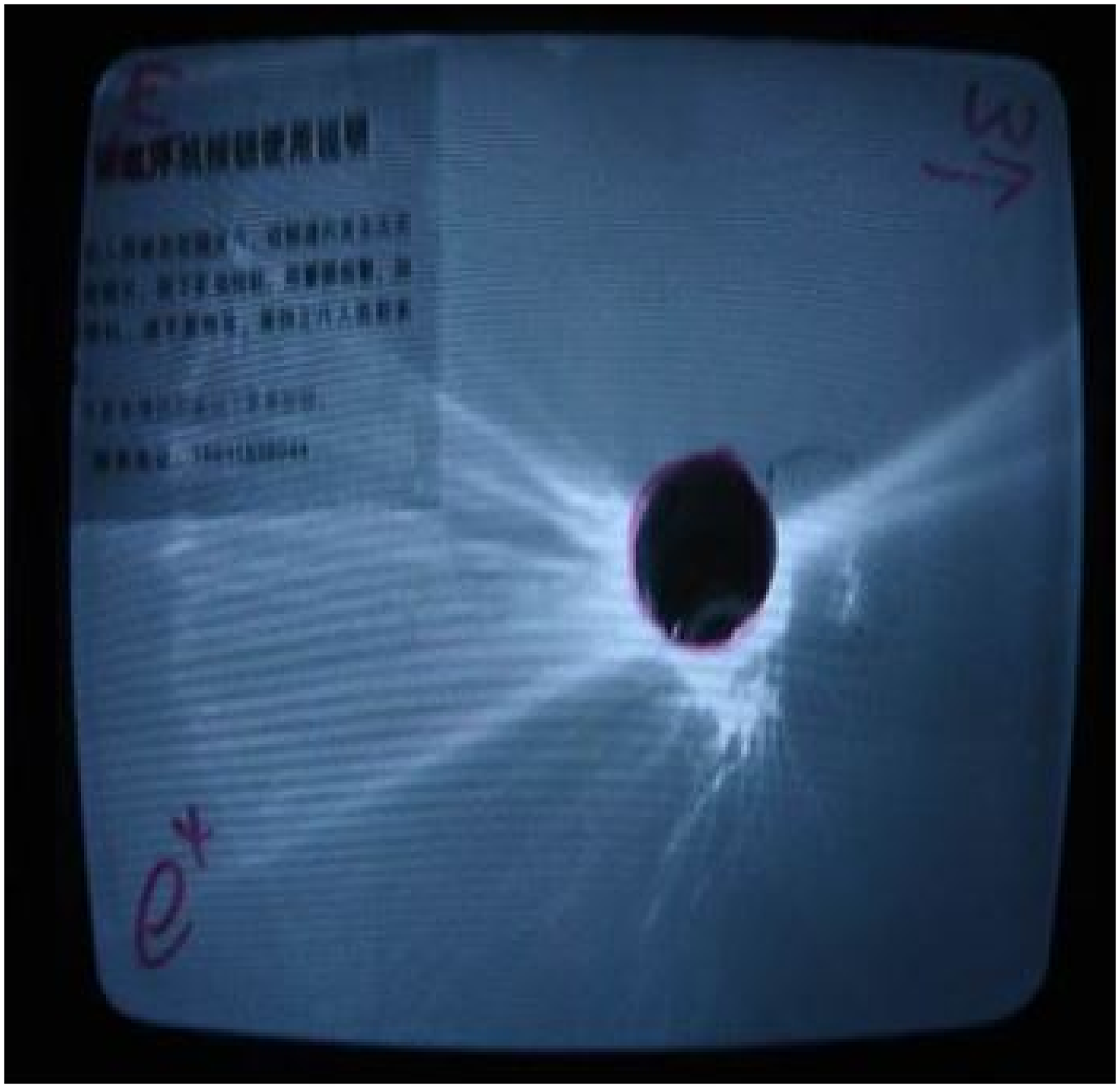}
\centerline{ (d)  }
\end{minipage}
\end{center}
\caption{\label{mirror} (a) The copper mirror mounted on a special support by screws, the adjustment for horizontal direction and vertical direction are separated.
(b) The simplified schematic of tuning the optical path.
The solid and dashed line indicate the light path before and after (after and before) tuning.
The circles mean the holes in the shielding wall in the BEPCII tunnel. By means of adjusting the direction of mirror, the light was moving into the holes in the wall.
(c) and (d) are the effect of the copper mirror adjustment using cameras. The white spots are the
synchrotron radiation and the black circles are the holes of wall in BEPCII tunnel.
(c) Before the copper mirror adjustment, the synchrotron light is reflected on the wall. (d) After the mirror adjustment (b), the synchrotron light is reflected into
the holes of the wall.
 }

In the light of Fig.~\ref{ln2fill}, it can be noted that the beam current of positron or electron is bent at the NCP. The collision between laser and current happens just before the bend, which avoids the interference of beam current with backscattering photons. At the same time, the synchrotron light radiates along the tangent direction of the bend beam and the path of this synchrotron light is the same path as that of the laser beam but with the opposite direction (refer to Fig.~\ref{ltov}). Therefore, the synchrotron light just pin down the laser beam alignment issue. The actual process of light path adjustment is expounded as follows.

The copper mirror is mounted on a special copper support, which can be turned by bending the vacuum flexible bellow using screwdriver. The back penal of the copper support is shown schematically in Fig.~\ref{mirror}(a), where the positions of four screws are denoted. The adjustment for the horizontal and the vertical directions are realized by two pairs of screws separately. The adjustment distance $l$ can be calculated as follows when the screw rotates one circle:
\begin{equation}
\frac{d}{D} = \frac{l}{L}~.
\label{calcult}
\end{equation}
where $d=0.5$ mm is the one circle distance for screw; $D$ is the dimension of the support, it is equal to 70 mm for the vertical support, to 39 mm for the horizontal one; $L=4150$ mm is the distance between the copper mirror and the hole in the wall.
During the running period of BEPC-II, the synchrotron light spots are visible. Two cameras, the east is for positron and the west for electron, are installed before the holes in the tunnel wall. The distance between the spot and the center of the hole is recorded. Comparing with the one circle distance obtained using Eq.~(\ref{calcult}), the number of turns can be evaluated, the optical path would be adjusted accordingly. Two pairs of screws accommodate two dimensional adjustment of light path. The cartoon drawn in Fig.~\ref{mirror}(b) shows schematically the tuning of the optical path. Usually one or two times tuning is enough. The actual adjustment effects by means of the cameras are displayed in Fig.~\ref{mirror} (c) and (d), by virtue of which it is clear to see that after the mirror adjustment, the synchrotron light passes through into the holes in the wall and is reflected to the laser output. On the contrary, the laser can be transferred through the hole and reflected into the vacuum pipe to make collision with the current beam.

The second advanced technique employed in this part is the craftmanship for making the viewport.
As shown in Fig.~\ref{ltov}, the entrance viewport is crucial for the laser to vacuum insertion system, since it is concerned to the light path adjustment and the laser beam insertion. Two things must be settled. The first one is that the viewport must be transparent to both laser beam and synchrotron light. Two types of entrance viewports~\cite{vpt,acha} are sequentially used by BEMS, one is based on gallium arsenide (GaAs) monocrystal plate and the other one is zinc selenide (ZnSe) polycrystal plate. The second thing is the connection between the stainless steel flange and the crystal plate. Here a special technique was developed to fixed the issue. As a matter of fact, in order to compensate mechanically the difference of the GaAs crystal and stainless steel thermal expansion coefficients, GaAs plate is brazing with pure quite soft lead to titanium ring, which in turn is brazing with AgCu alloy to the stainless steel ring. The stainless steel ring is weld to the flange.

At the beginning, the GaAs plate with diameter of 50.8 mm and thickness of 3 mm is adopted by BEMS. However,  the GaAs crystal is not transparent for the visible light, but it transmits the infrared radiation. In order to detect the spots of infrared light, IR-sensitive video cameras were used. During the BEMS running, the kind of viewport has been changed for three times. Finally the ZnSe plate with  thickness of 8 mm is adopted. The laser transmission rate goes up from 60\% to 76\%. In addition, more visible synchrotron light is transparent, which is convenient for the optical path adjustment.

\section{HPGe detection system}\label{sxt:hpgedtr}
As aforementioned in the working scheme, the backscattering high energy photons will be detected by the HPGe detector, which is the key instrument of BEMS. The accuracy of beam energy depends solely on the detection results of the HPGe detector. There are two crucial conditions for HPGe detector functioning properly and enduringly for BEMS at BEPC-II, that is low temperature and radiation protection. The upgradations about these two aspects are depicted below.

\subsection{Cryogenic system upgrading}
\subsubsection{HPGe detector}
The purpose of a HPGe detector is to convert gamma rays into electrical impulses to determine their energy and intensity.

A p-type coaxial HPGe detector manufactured by ORTEC (model GEM25P4-70) is adopted by BEMS , whose energy resolution for the 1.33 MeV peak of
$^{60}$Co is 1.74 keV, with the relative efficiency is 25$\%$. The detector is connected to the multichannel analyzer of ORTEC DSpec Pro(MCA), which transfers data using the USB port of the computer.

\subsubsection{Two cryogenic systems}
Low temperature is crucial for HPGe detector functioning properly~\cite{knoll,cooleruser,coolerii}. Two approaches are usually employed to get the temperature below 100 kelvins (K) : liquid nitrogen (LN$_{2}$) and electric coolers~\cite{hpge}. The former is firstly utilized for cooling the HPGe detector at BEMS. The germanium crystal is cooled down to the working temperature by connecting with the thermal transfer device, the cryostat and the extension rod that are dipped into the LN$_{2}$-full dewar, and the cold is conducted from the dewar to the germanium crystal.

One common filling method use a standard 30-liter dewar of LN$_{2}$ as a supply dewar, a so-called self-pressurizing technique is employed for LN$_{2}$ filling, more details can be found in Ref.~\cite{info}. During the data taking period of BES-III, the BEMS is kept running simultaneously. The once-a-week filling schedule is followed to avoid unexpected warm-ups of HPGe detector.  However, such a regular filling schedule is unfavorable to both BES-III detector and BEPC-II accelerator. For the detector, some precious data taking time has to be consumed for refilling LN$_{2}$; for the accelerator, some time has to used to recover the preceding good running status.

From the point view of continuous cooling, a electric cooler is an ideal replacement for the LN$_{2}$. The electric cooler, composed of a compressor, transfer hose, heat exchanger, and cold head, is adopted for HPGe detector. The only concern here is the continuous electricity power, therefore a uninterruptible power system (UPS) is used for the cooler, which guarantees the power supply.

\subsubsection{Resolution comparison of two cryogenic systems}
As a matter of fact, the electric cooler was once the first choice of BEMS, but it was a concern on the resolution of HPGe under such cryogenic system. Since the range of
cooling temperature for electric cooler is from 85 to 105 K that is a little bit
higher than that for LN$_{2}$ (the boiling temperature for LN$_{2}$ is 77 K
at standard temperature and pressure), it is believed that the resolution under liquid cooling condition would be better. A laboratory experiment was designed and performed~\cite{info} to investigate the resolutions under
the two cryogenic conditions.

During the experiment, a point like radiation source of $^{152}$ Eu, whose main lines are from hundred keV to 1.4 MeV\cite{eu1,eu2}, is placed along the cylindrical center axis of the germanium crystal, and about 1 cm far away from the top of the germanium detector. A 1 cm foam plate is inserted between the source and the detector. The HPGe detector is calibrated by $^{137}$Cs and $^{60}$Co before experiment.

The experiment began with the electric cooler case.
About three days' data were collected when the HPGe exposed to radiation source. In order to remove the background effect, 3 days' background data were taken before and after the $^{152}$Eu nuclide measurement, separately.

\begin{center}
\includegraphics[height=5.0cm]{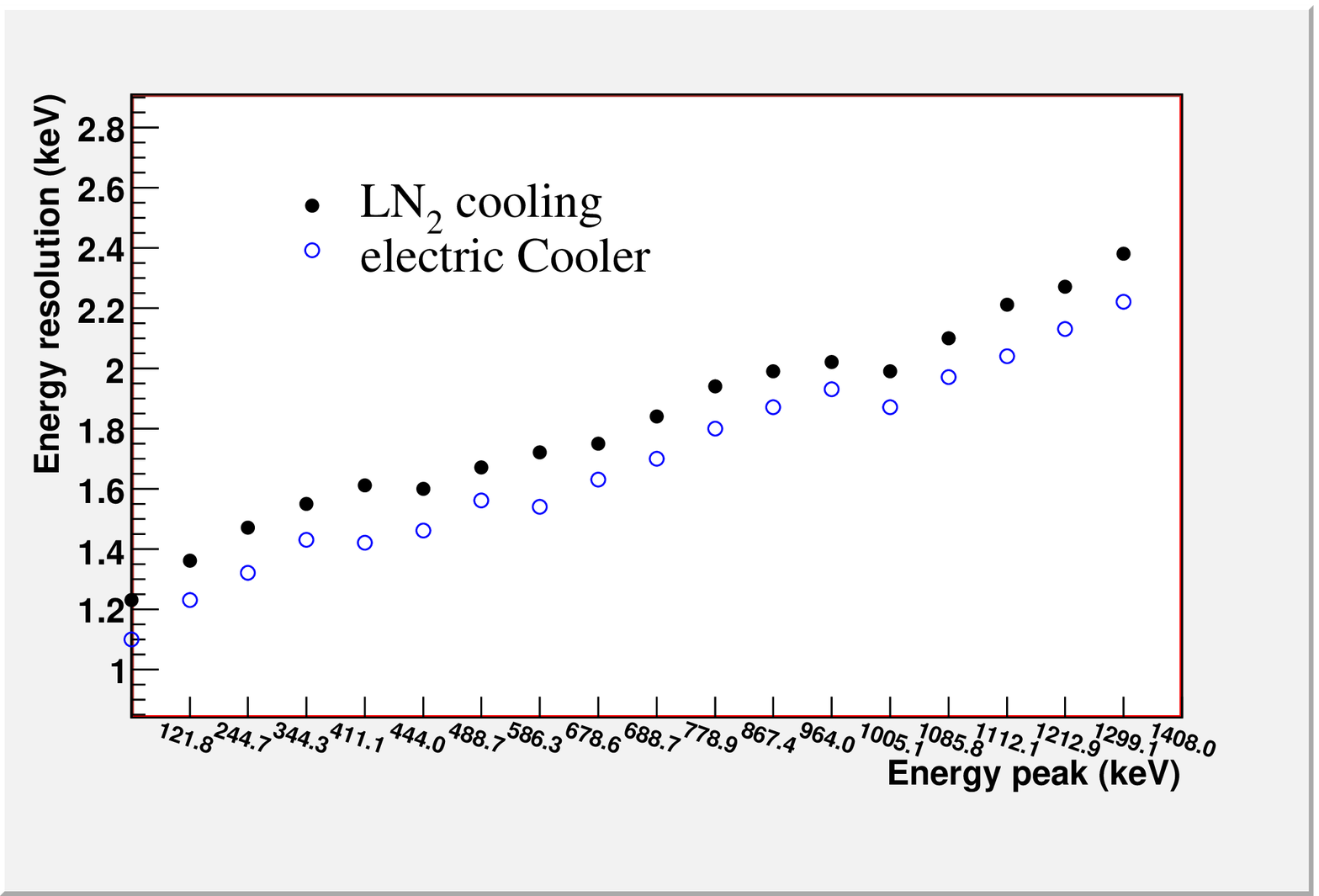}
\caption{\label{cooler}
The comparison of the detector resolutions to the characteristic lines of $^{152}Eu$ under $LN_{2}$ cooling method represent as dot and electric cooler method indicated as circle.}
\end{center}

After above experiment, the PopTop capsule of the detector
was removed from the cold head of the electrical cooler, then connected with the cryostat,
and put into a dewar filled with LN$_{2}$. After about 6 hours cooling, the germanium crystal is cold enough to
apply the high voltage to bias the detector.
The detector was calibrated using $^{137}$Cs and $^{60}$Co before $^{152}$Eu measurement under LN$_{2}$ cooling condition.
The radiation source experiment was performed about three days.
Also the background data of about 3 days were taken before and after the measurement of $^{152}$Eu under the LN$_{2}$ cooling, respectively.

The comparison of the detector resolutions to the characteristic lines of $^{152}$Eu under different cooling methods are shown in Fig.~\ref{cooler}. As depicted in Ref.~\cite{info}, the shape of the lines are almost same, but the resolution of the germanium detector using the electric cooler is
about 10\% better than using the LN$_{2}$ case. The noise level for both LN$_{2}$
and electric cooler is same, about 10 keV.

The laboratory measurements indicate that the resolution of HPGe detector under the electrical cooler is better than that of liquid nitrogen cooling condition. Therefore, the electric cooler is installed in summer of 2013 to replaced the $LN_{2}$ cooling.

\subsection{Alternating moving shielding}
Since the HPGe detector is located near the beam pipes of collider, radiation background due to beam loss is extremely high~\cite{bkgd1,bkgd2}. In order to protect the HPGe detector from radiation damage, the special design of radiation protect is indispensable~\cite{bkgd1}. In actual running period, the detector is surrounded by 5 cm of lead on the sides, by 1.5 cm of iron below, and by 5 cm of lead above. Moreover, it is also shielded by 10 cm of paraffin on all sides. Since the main radiation background comes from the beam direction, an additional 11 cm of lead is installed in these directions~\cite{bems}.

However, even with above protections, the radiation background along the beam direction seems still high. For improvement, an alternating moving shielding device is designed for the further protection. As shown in Fig.~\ref{shield} (a), two movable stages with 10 cm thickness of leads are fixed on the Aluminum electric push rod, they can move in the range of 350 mm and its movement speed is 8 mm per second. The electric push rod is installed between the HPGe detector and the short vacuum chamber.
If need, these leads can move into the beam direction to shield the high energy photons coming from the non-measurement direction. For example, assuming the energy of positron beam need to be measured, the movable lead at the positron side (east side) will move out from the beam direction, and the backscattered photon will enter into the sensitive volume of germanium detector for measurement. However the electron side (west side) movable lead will move into the direction to shield the radiation photons from the election beam. The working flow chart of the movable shielding is shown in Fig.\ref{shield} (b).
\begin{center}
\begin{minipage}{5cm}
\includegraphics[height=4.0cm,width=6.cm]{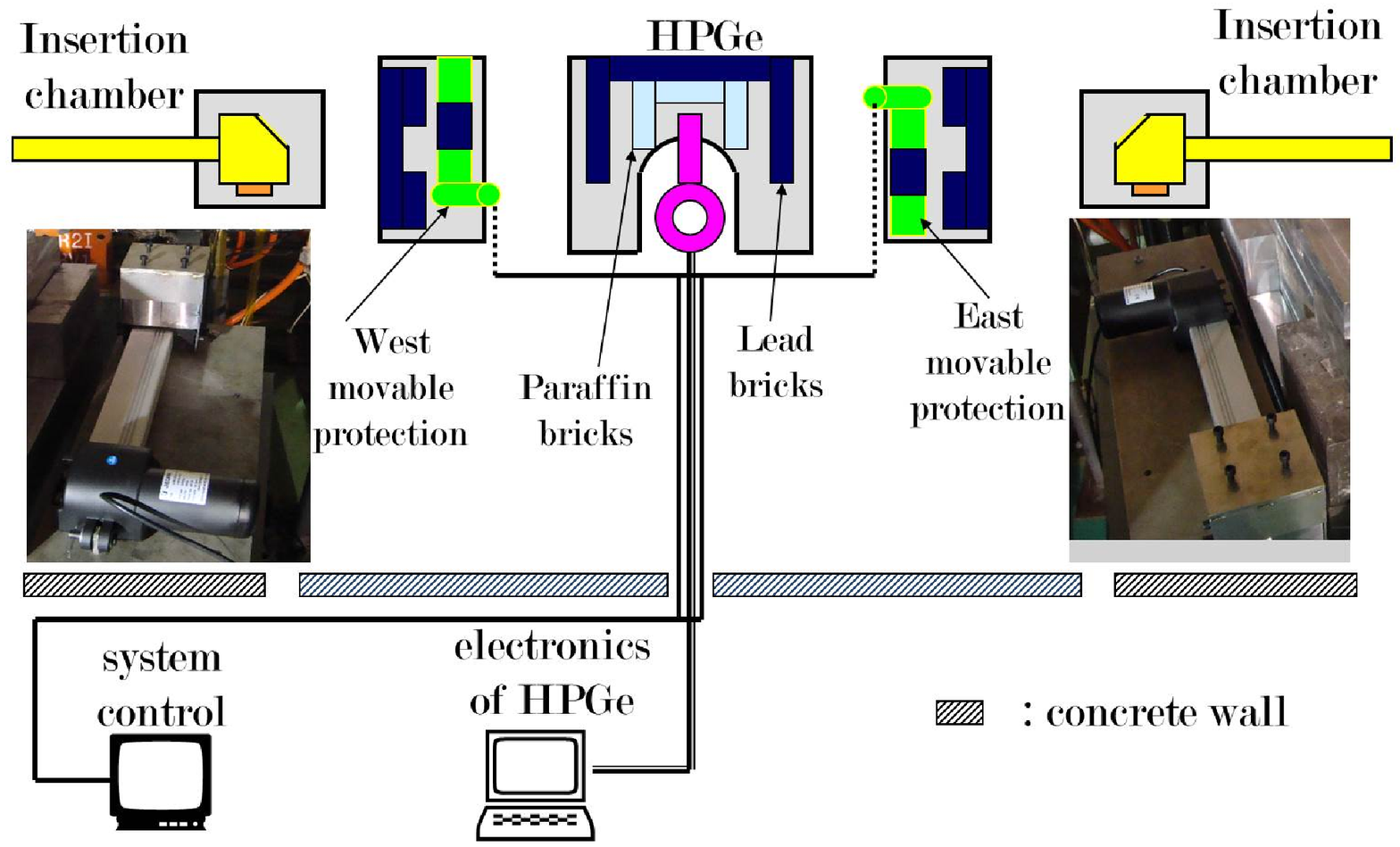}
\centerline{ (a)}
\end{minipage}
\begin{minipage}{4cm}
\includegraphics[height=4.0cm,width=4.cm]{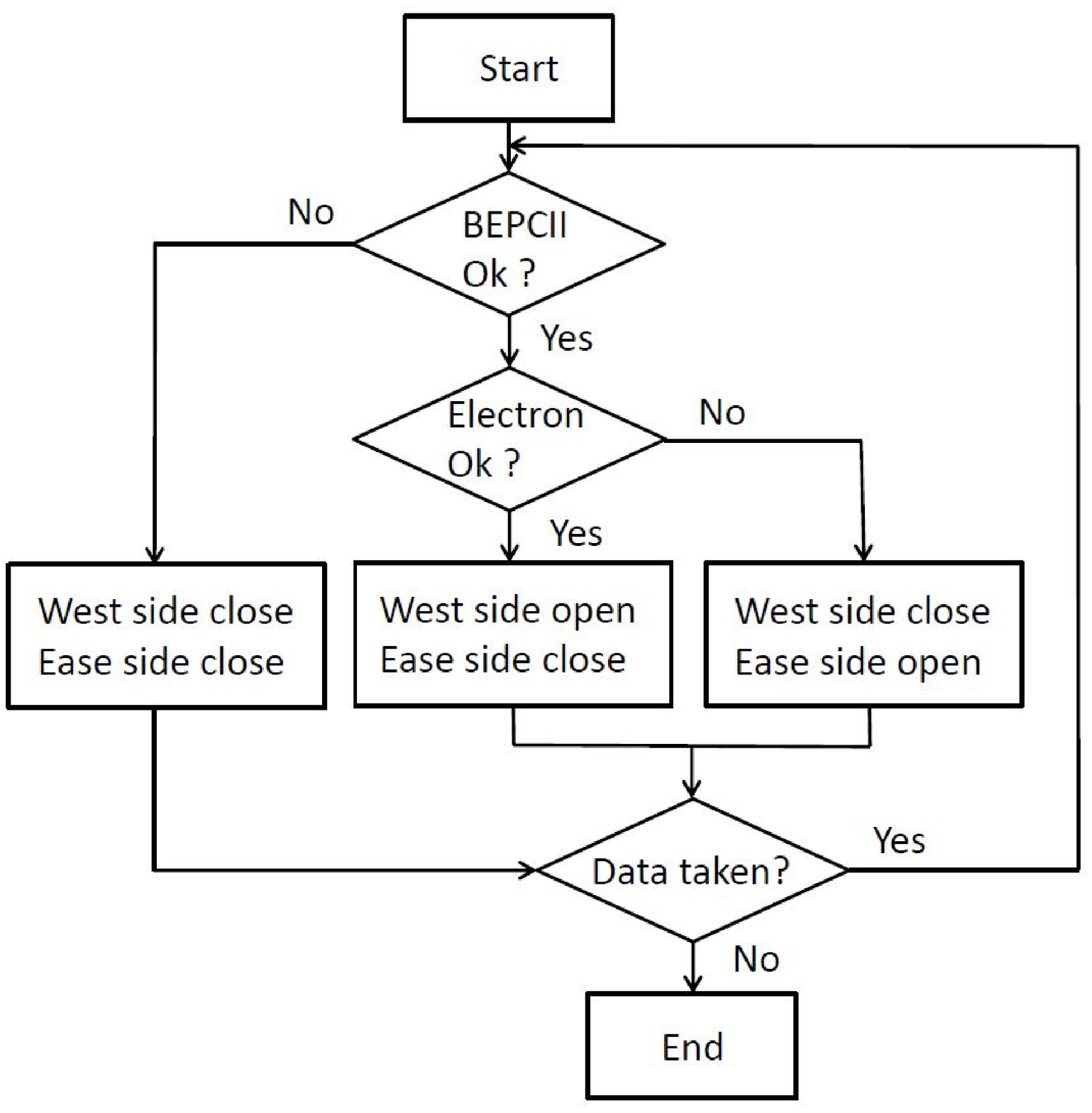}
\centerline{ (b)}
\end{minipage}
\begin{minipage}{4cm}
\includegraphics[height=3.5cm]{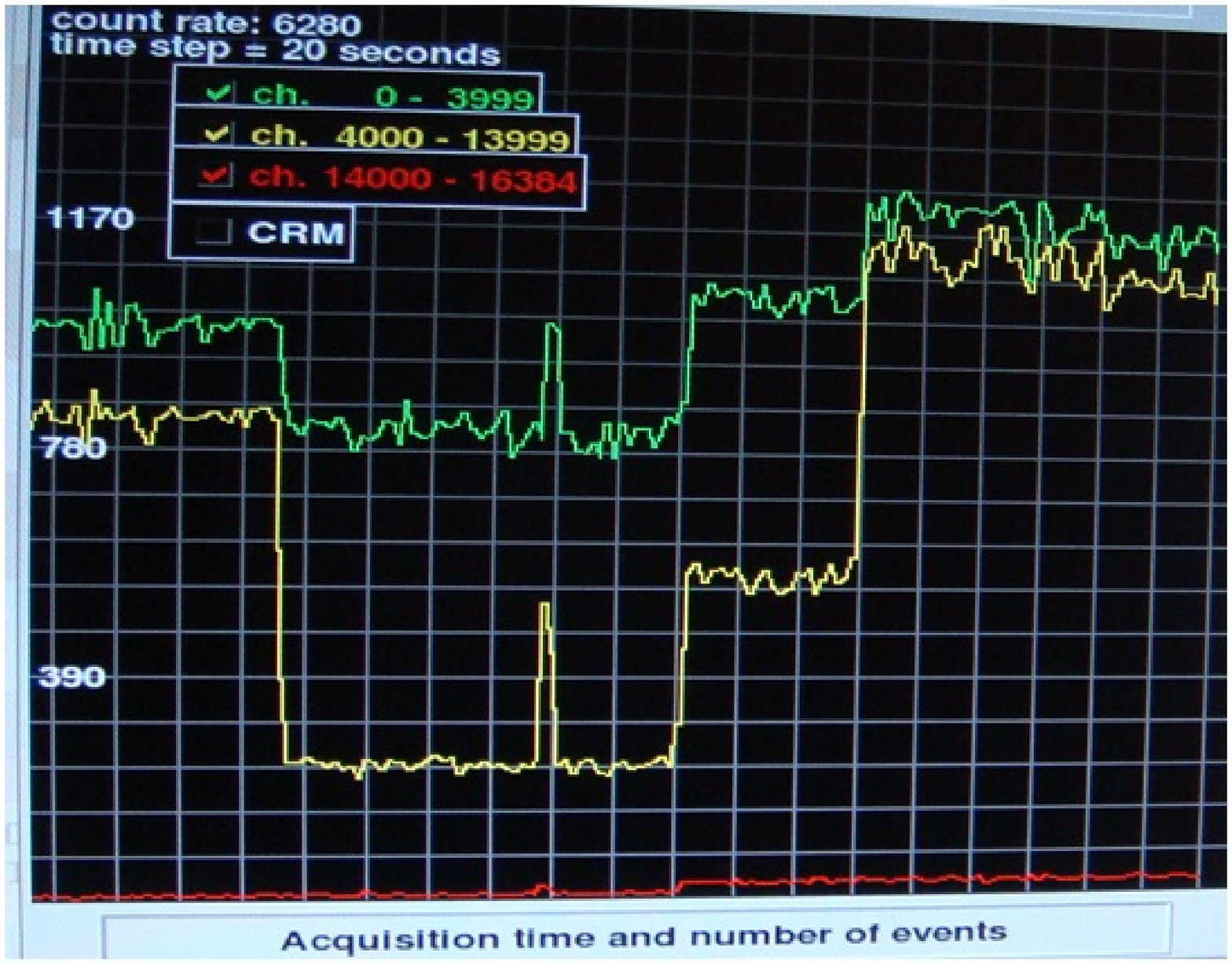}
\centerline{ (c)}
\end{minipage}
\caption{\label{shield}The movable shielding. (a) The sketch of the movable shielding, which is installed between the HPGe detector and laser insertion chamber.
(b) The flow chart of the movable protection working principle. (c) The
improvement of the signal to background ratio using moving shielding.
The green line presents the low energy background, the yellow line illustrates the
 (S + B). when the laser is turned off, the yellow line illustrates the pure background (B).
The left part shows the case that the moving shielding is used, the S/B is about 3.92, without out moving shielding case is shown in the right part, the S/B is about 2.08. }
\end{center}

The improvement due to the shielding device for the signal to background ratio is shown in Fig.~\ref{shield}(c), from which it can be seen that one fold increase of the signal to background ratio is obtained by virtue of the alternating moving shielding device.

\section{Data acquisition and processing}
\subsection{Data acquisition system}
The relation between the maximal energy of the scattered photon $\omega_{max}$ and the beam energy $\varepsilon$ is provided by Comptom scattering theory~\cite{rus, moxh}£¬ {\rm i.e.}
\begin{equation}
\omega_{max} = \frac{\varepsilon^{2}}{\varepsilon + m^{2}_{e}/4\omega_{0}}~.
\end{equation}
Here $\omega_{max}$ can be determined through the detection of scattered photons by HPGe detector. Then the beam energy can be deduced from above formula,
\begin{equation}
\varepsilon = \frac{\omega_{max}}{2} \left[1+\sqrt{1+\frac{m^{2}_{e}}{\omega_{0}\omega_{max}}}\right]~.
\label{nepec}
\end{equation}
The actual data acquisition system is finished automatically, which controlled by a software. Its
working procedure is as follows:  Firstly, some requirements (such as data taken time, data type, or energy difference range)
are input as parameters into the software, then the software visits the BEPCII database and gets the status parameters of the accelerator, such as beam currents, lifetime, energy value and so on. Then the HPGe detector begins to take data. Every about few seconds, the HPGe detector measurements are stored and the detector counting rate is calculated. The mirrors are adjusted automatically by a maximal photon/beam interaction efficiency using the feedback from the detector counting rate.

If the status of the accelerator is changed sufficiently, such as energy drift or beam loss, the current spectrum is saved and named as the end time of data taken, the next spectrum acquisition cycle also at the same beam side is launched. Simultaneously, another program processes the saved spectrum, calibrates the energy scale, finds the Compton edge, and calculates the beam energy. The beam energy is written into the BEPC-II database.

The measurement will switch to the other side of beam when the requested data acquisition time has finished or the status of measurement beam is not satisfy the requirement. The rotary stage will turn a certain degree and direct  the laser into the other side of the vacuum chamber for collision. The movable leads will move in or out the beam direction according to the requirement.
All these adjustments are finished automatically.

\begin{center}
\begin{minipage}{4cm}
\includegraphics[height=4.0cm,width=6.0cm]{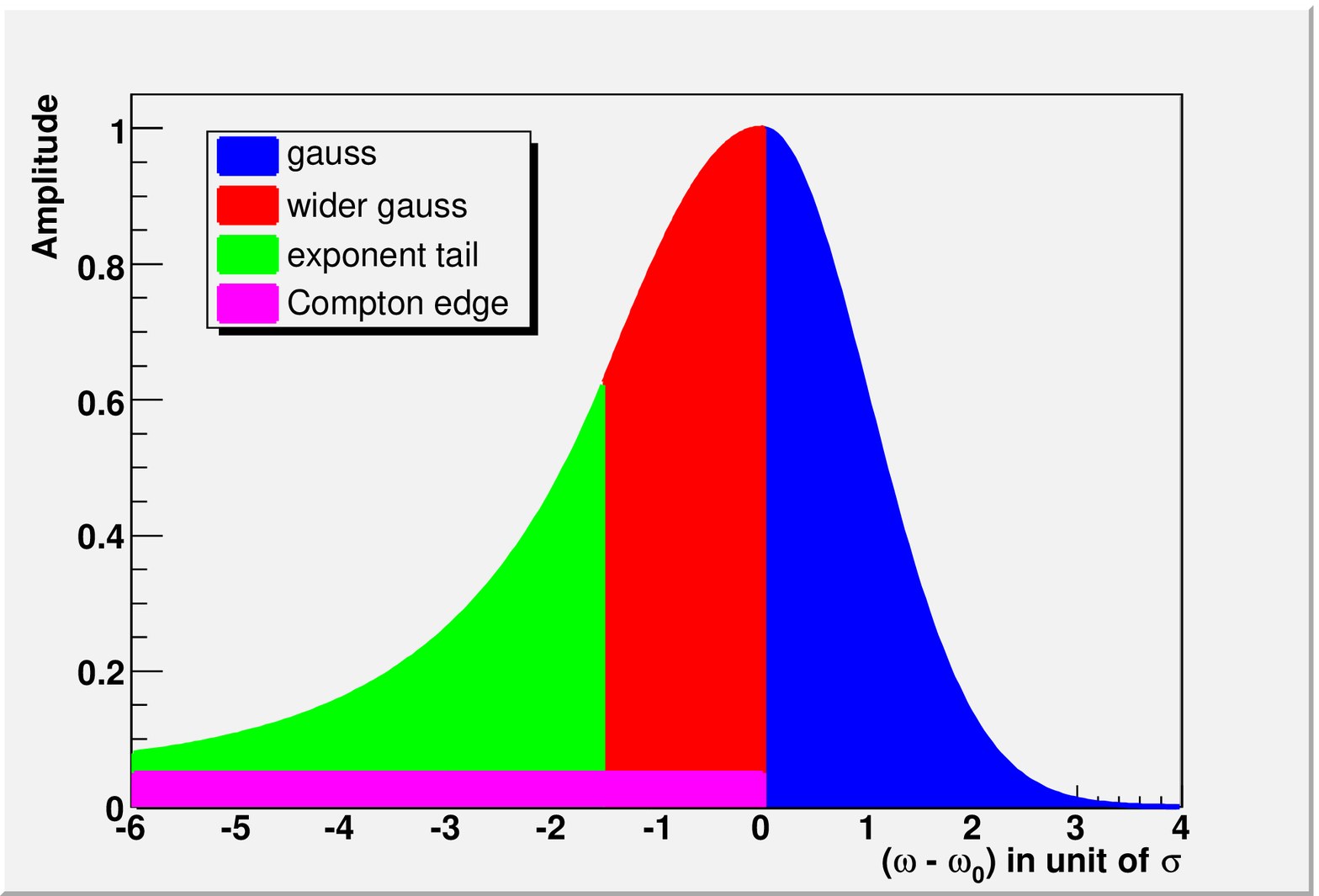}
\end{minipage}
\caption{\label{dasandgauss}
The improved response function of the HPGe detector.}
\end{center}

\subsection{Calibration improvement}
The kernel of data acquisition system lies in the data processing, which is composed of three parts, that is calibration of energy scale, Compton edge fitting, and determination of beam energy.
The improvement mainly consists in the response function and calibration source.

\subsubsection{Response function and edge fitting}
The goal of calibration is to obtain the coefficients needed for conversion of detector's ADC counts into corresponding energy deposition, measured in units of keV, as well as determination of the detector's response function parameters. The following response function was used:


\begin{equation}
 f(x,x_0,\sigma,\xi) = A \cdot
\left\{\begin{array}{ll}
  \exp\biggl\{ {-{(x-x_0)^2\over 2\sigma^2}} \biggr\},
  & x > x_0-\xi\cdot\sigma \\
  \exp\biggl\{ {\xi^2\over 2}{+{\xi(x-x_0)\over \sigma}} \biggr\},
  & x < x_0-\xi\cdot\sigma,  \\
\end{array}\right.
\label{m-gauss-norm}
\end{equation}
where A is amplitude with normalization, $x_0$ is the position of peak, $\xi$ is an asymmetry parameter, and $\sigma$ is the full width of gaussian distribution at half maxium divided by 2.36.

The edge of backscattered photons spectrum is fitted by the function:
\begin{equation}
S_2(x,x_0,\sigma,\sigma_s,\xi) = \int\limits_{x}^{+\infty}
S_1(y,x_0,\sigma,\sigma_s,\xi)\;dy\; +p_1(x).
\label{convolution-5}
\end{equation}
Here $p_1(x)$ takes into account background contribution and $S_1$ is a convolution of the step function $\theta(x_0-x)$:
\begin{equation}
\theta(x_0-x) =
\left\{\begin{array}{ll}
  1,  & x < x_0  \\
  0, &  x > x_0,  \\
\end{array}\right.
\end{equation}
which describe the ``pure'' edge shape with HPGe detector response function
(\ref{m-gauss-norm}) and gaussian:
\begin{equation}
 g(x,x_0,\sigma_s) = {1\over \sqrt{2\pi}\sigma_s}
 \exp\biggl\{ -{{(x-x_0)^2}\over{2\sigma_s^2}}\biggr\},
\end{equation}
which takes into account the energy spread of backscattered photons due to energy distribution of the collider beam.
\begin{eqnarray}
\nonumber & S_1(x,x_0,\sigma,\sigma_s,\xi) =
\displaystyle\frac{N}{2\sqrt{2\pi}} \times &  \\
& \times \Biggl[  \displaystyle\frac{1}{\sigma}
\exp\Bigl(\frac{\xi^2}{2}\bigl(1+\frac{\sigma_s^2}{\sigma^2}\bigr)+
\frac{\xi x}{\sigma}\Bigr)
\cdot\mathrm{erfc}\Bigl(\frac{\xi(\sigma^2+\sigma_s^2)+x\sigma}
{\sqrt{2}\sigma\sigma_s}\Bigr) + &\\
\nonumber & +  \displaystyle\frac{1}{\sqrt{\sigma^2+\sigma_s^2}}
\exp\Bigl(-\frac{x^2}{2(\sigma^2+\sigma_s^2)}\Bigr)
\cdot\mathrm{erfc}\Bigl(-\frac{\xi(\sigma^2+\sigma_s^2)+x\sigma}
{\sqrt{2(\sigma^2+\sigma_s^2)}\sigma_s}\Bigr) \Biggr] & .
\label{convolution-4}
\end{eqnarray}
The edge position $\omega_{max}\equiv x_0$, $\sigma_s$ and coefficients of the
first-order polynomial $p_1(x)$ are the free parameters of fit.
Using the $\omega_{max}$ value obtained from fit, the average beam energy
$\varepsilon_{nip}$ ($nip$ denotes north interaction point) in the $e-\gamma$
interaction region is calculated according to formula (\ref{nepec}).
Taking into account the energy losses due to synchrotron radiation, the beam
energy in the south interaction point ($sip$) is obtained as
\begin{equation}
 \varepsilon_{sip}(\mathrm{MeV})  = \varepsilon_{nip}(\mathrm{MeV}) +
 4.75\cdot10^{-3}*(0.001 \cdot \varepsilon_{nip}(\mathrm{MeV}))^4.
\label{cme}
\end{equation}

The above response function, formula (\ref{m-gauss-norm}), can not depict the distribution of characteristic peak of
radiation source precisely, a wider gaussian is introduced to improve the description of the left part of the peak.
the new response function, as shown in Fig.\ref{dasandgauss}, becomes complex as follows:
\end{multicols}
\begin{equation}
f(x)=A \cdot \left\{
\begin{array}{cc}
\displaystyle exp[-\frac{(x-x_{0})^{2}}{2\sigma^{2}}],  \ \ & \ \  0 < x-x_{0} < + \infty,\\
\displaystyle C+(1-C)exp[- \frac{(x-x_0)^{2}}{2(K_{0}\sigma)^{2}}],   \ \ & \ \ -K_{0}K_{1}\sigma <x-x_{0} \leq 0,\\
\displaystyle C+(1-C)exp[K_{1}(\frac{(x-x_0)}{(K_{0}\sigma)} + \frac{K_{1}}{2})],   \ \ & \ \ - \infty < x-x_{0} \leq -K_{0}K_{1}\sigma,
\end{array}
\right.
\label{m-gauss-norm-new}
\end{equation}
\begin{multicols}{2}
where $K_{0}\sigma$ is the deviation of the wider gaussian distribution from $x_0 - K_{1}(K_0\sigma)$ to $x_0$, $x_0 - K_{1}(K_0\sigma)$ is the position where the exponential tail appears. C is responsible for the small angle Compton scattering of $\gamma$ quanta in the passive material between the source and the detector. $K_{1}$ is the asymmetry parameter.

The calibration procedure is same as mentioned before. The parameters $x_0$, $\sigma$, $K_{0}$, $K_{1}$, $C$ are determined when the calibration peaks are identified and fit.

After calibration, we need to measure the energy position of the sharp edge of the energy spectrum of backscattered photons. The function to fit the edge was obtained in two steps. Firstly, we calculate the convolution of the response function with another gaussion, which is responsible for the energy spread in the beam:
\end{multicols}
\begin{equation}
\begin{array}{cc}
\displaystyle S(x) = \frac {1}{2\sqrt{2\pi}}\{ \frac{1}{\sqrt{\sigma^{2}+\sigma_{s}^{2}}} exp(\frac{-x^{2}}{2(\sigma^{2}+\sigma_{s}^{2})})erfc(\frac{-x\sigma}{\sigma_{s}\sqrt{2(\sigma^{2}+\sigma_{s}^{2})}})
\displaystyle +\frac{C}{\sigma}erfc(\frac{x}{\sqrt{2}\sigma_{s}})\\
\displaystyle + \frac{(1-C)K_{0}}{\sqrt{\sigma_{m}^{2}+\sigma_{s}^{2}}}
\displaystyle exp(\frac{-x^{2}}{2(\sigma_{m}^{2}+\sigma_{s}^{2})})(erfc(\frac{x\sigma_{m}}{\sigma_{s}\sqrt{2(\sigma_{m}^{2}+\sigma_{s}^{2})}})-
\displaystyle erf(\frac{K_{1}(\sigma_{m}^{2}+\sigma_{s}^{2})+x\sigma_{m}}{\sigma_{s}\sqrt{2(\sigma_{m}^{2}+\sigma_{s}^{2})}}))\\
\displaystyle +\frac{1-C}{\sigma}exp(\frac{K_{1}^{2}}{2}(1+\frac{\sigma_{s}^{2}}{\sigma_{m}^{2}})+K_1\frac{x}{\sigma_{m}})
\displaystyle erfc(\frac{K_{1}(\sigma_{m}^{2}+\sigma_{s}^{2})+x\sigma_{m}}{\sqrt{2}\sigma_{s}\sigma_{m}})
\}.
\end{array}
\end{equation}
\begin{multicols}{2}
Here $x$ means $x-x_{0}$, the $\sigma_{s}$ is responsible for the beam energy spread, $\sigma_{m} = K_{0}\sigma$.

Then, the edge is fitted with:

\begin{equation}
S_{1}(x,x_{0},\sigma,K_{0},K_{1},C,\sigma_{s}) = B + \int\limits_{x}^{+\infty}S(y,x_{0},\sigma,K_{0},K_{1},C,\sigma_{s})dy.
\end{equation}
Where B is the background. During edge fitting, parameters $\sigma$, $K_{0}$, $K_{1}$ are fixed to the values obtained from the calibration. From the fitting results, one can get the beam energy, energy spread, and the background.

\subsubsection{Calibration source upgrade}

The energy to be measured at BEPC-II ranges from 1 to 2 GeV that corresponds the energy range of backscattered photons from 2 to 6 MeV. The calibration sources used from the very beginning of BEMS commissioning are as follows:
\begin{itemize}
\item
 $^{137}$Cs      :  $E_\gamma = 661.657  \pm 0.003$~keV
\item
 $^{60}$Co       :  $E_\gamma = 1173.228 \pm 0.003$~keV
\item
 $^{60}$Co       :  $E_\gamma = 1332.492 \pm 0.004$~keV
\item
$^{16}$O$^*$    :  $E_\gamma = 6129.266 \pm 0.054$~keV
\end{itemize}
However, between 2 and 6 MeV, there is no source for calibration. To remedy such a defect, besides radiation sources listed above, a precise pulse generator is adopted for detailed calibration purpose.

The ORTEC$^{\textcircled{\scriptsize R}}$ DSPEC Pro$^{\mbox{TM}}$ MCA has a $\pm$250 ppm declared integral non-linearity (in 99.5\% of scale). A precision pulse generator (BNC model PB-5) is declared by manufacturer to have $\pm$15 ppm integral non-linearity and $\pm$10 ppm amplitude jitter. The PB-5 outlet is connected to HPGe preamp inlet to provide a set of discrete pulse amplitudes, forming a corresponding set of peaks in the measured spectrum. The PB-5 pulse shaping parameters were selected
as follows:
\begin{itemize}
\item
Attenuation: 10 ,
\item
Pulse rise time: 50 ns (minimum) ,
\item
Pulse width: 14 $\mu$s (pulse top: flat) ,
\item
Pulse fall time: 500 $\mu$ ,
\item
Pulse amplitudes: {0.75, 1.20, 1.40, 1.65, 1.75, 2.10, 2.50, 2.90, 3.40, 4.00, 4.50, 5.00, 5.40, 6.10, 6.50, 7.00, 8.50, 9.00, 10.0} V, forming 19 calibration peaks ;
\item
Pulse repetition rate: 20-50 Hz .
\end{itemize}
Switching between different amplitudes listed above occurs randomly in time via a simple computer script.

To test the calibration effect, two reference lines from [$^{232}$Pu~$^{13}$C] gamma are used.
$\alpha$-decay of $^{232}$Pu provides the reaction:
\begin{equation}
 \alpha + ^{13}\!\mbox{C} \rightarrow n + ^{16}\!\mbox{O}^*~~.
\label{eq:PuC}
\end{equation}
An exited oxygen nuclei emits $\gamma$-rays with the energy of 6129.266 $\pm$ 0.054~keV~\cite{Alkemade1982383}.
The p-type HPGe detector used at BEPC-II is shielded from neutrons, emitted in the reaction (\ref{eq:PuC}), by about 10 cm of paraffin. The presence of these neutrons lead to the reaction:
\begin{equation}\label{eq:npdg} n + p \rightarrow d + \gamma~~,\end{equation}
from which we can observe the 2223 keV $\gamma$-rays as by-product of such a configuration. This energy can be found in Refs.~\cite{Kessler,Dataofgamma2007} to be $2223.24835\pm 0.00008$ keV.

\end{multicols}
\tabcaption{\label{cmpofsource}The comparison of measured energies from $\gamma$-ray and pulser generator.}
\begin{center}
\begin{tabular}{llll} \hline \hline
Energy    & $\gamma$-ray           & Pulser                & Reference  \\ \hline
6129 keV  & $6129.451\pm0.064$~keV &$6129.208\pm0.062$~keV &$6129.266 \pm 0.054$~keV \\
2223 keV  & $2223.144\pm0.272$~keV &$2223.149\pm0.022$~keV &$2223.24835\pm 0.00008$~keV  \\ \hline
\end{tabular}
\end{center}
\begin{multicols}{2}
Table~\ref{cmpofsource} lists the results of measured energies from $\gamma$-ray and pulse generator. The comparison confirms the reliability of pulser calibration.

\begin{center}
\includegraphics[width=8.0cm]{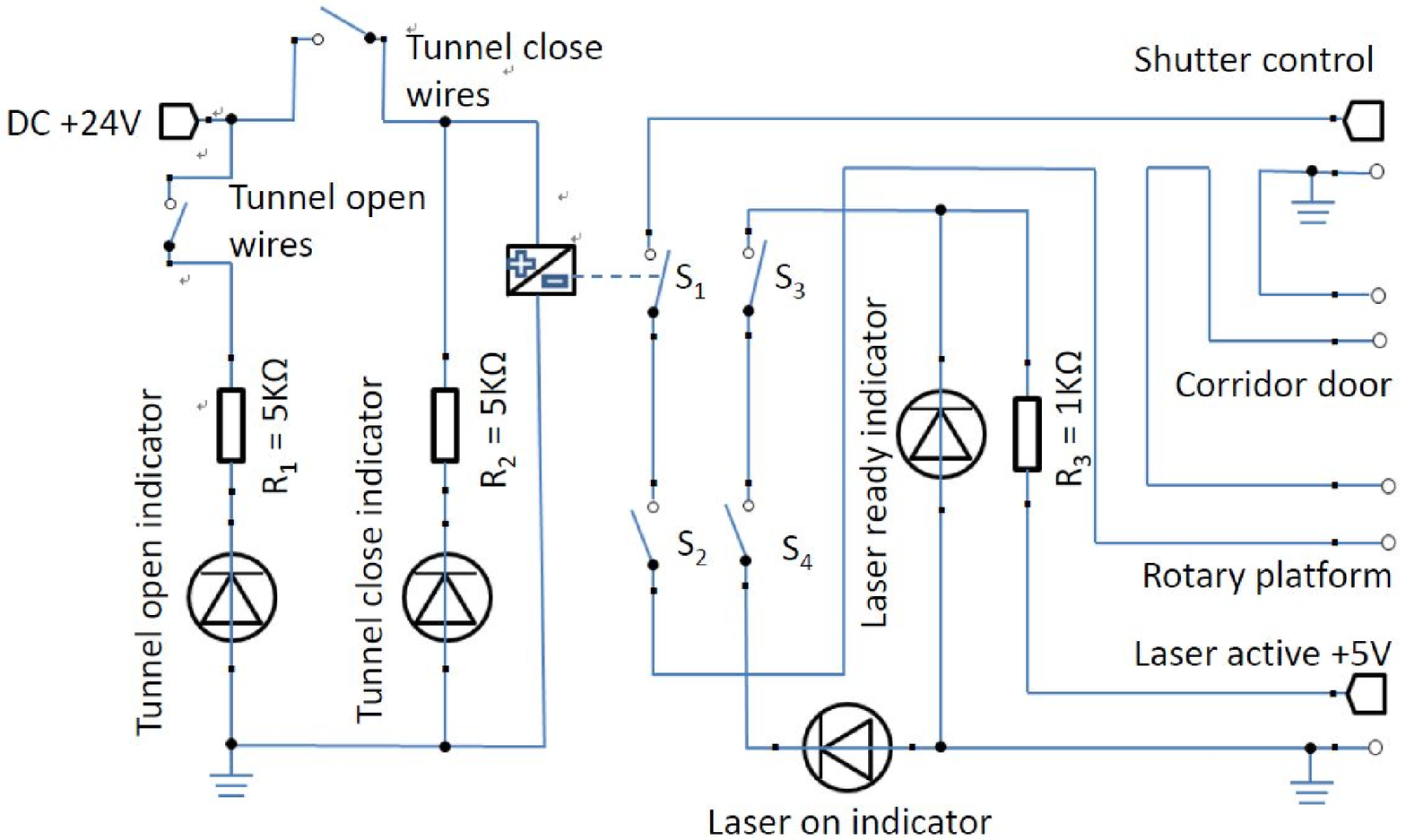}
\figcaption{\label{interlock} Schematic diagram of interlock of BEMS.}
\end{center}

\section{Laser interlock}

Laser is invisible, dangerous to person who works at corridor or tunnel near NCP of BEPC-II. Interlock is a good way to protect people from the laser damage. Figure~\ref{interlock} shows the schematic diagram of interlock of BEMS. Only when people evacuate from the BEPC-II storage ring,  and the tunnel door closed, the BEMS laser can be active, the indicator light will turn on.  After the shutter control switches on, the laser will be exported.

As mentioned before, the laser is located at the corridor. It will be dangerous to the BEMS staff who enter the corridor, therefore the door of corridor is interlocked with laser. When the corridor door is open, the laser output will be terminated automatically.

When the moving prism is replaced by the rotary platform, the beam measurement switch between electron and positron becomes dangerous, because the laser will rotate 180 degrees according to the platform rotary. Once laser meets the flammable material, it is likely to cause fire. Therefore,
the status of BEMS has to be interlocked with laser. During the beam measurement switch, the rotary platform is rotating, the laser will turn off.

A total switch is installed near the corridor gate, it is always disconnect during the BEMS elements test and adjust in the corridor. Only when all operations are performed, turn on the total switch, the BEMS laser will export.

\section{Summary}
\end{multicols}
\tabcaption{\label{up}The upgradation chronicle of BEMS}
\begin{center}
\small
\begin{tabular}{|c|c|c|c|} \hline \hline
 Previous system   & Upgraded system & Upgrade time & Improvement \\ \hline
 Focus lenses ZnSe  &new focus lenses       &2011.4& laser transmission rate rise to 98\%\\ \hline
 No moving shielding&10 cm moving shielding &2011.9& higher signal background ratio\\ \hline
 No interlock       & USB interface relay   &2012.3& protect staff from laser \\ \hline
 Moving prism       & rotary platform       &2012.3& more durable \\ \hline
 $LN_{2}$ cooling   &electric cooler cooling&2013.8& $LN_{2}$ refill time is saved \\ \hline
 GaAs viewport      & ZnSe viewport         &2014.8&laser transmission rate rise to 76\%\\ \hline
 Laser with line 10P42 & Laser with line 10P20 & 2014.12 & laser power rise to 50 W\\ \hline
\end{tabular}
\end{center}
\begin{multicols}{2}
The established BEMS greatly increases the measurement capacity for both BEPC-II accelerator and BES-III detector, and becomes the indispensable part of them. Many technique details are provided for clearly  understanding the working process of the whole system; especially the upgrade improvement is expound. Listed in Table~\ref{up} are the mainly upgraded components of BEMS during the past several years. In which more powerful laser, higher transmission rate of ZnSe lenses and viewports will help to shorten the data taken time for one measurement. The laser accelerator interlock will prevent people from damage. The usage of electric cooler will help to save the precious running time from LN$_{2}$ refilling. New calibration response function for the HPGe detector and fitting
function for the energy sprectrum will help to get more precise beam energy.

\end{multicols}
\tabcaption{\label{info}The typical values of beam energy and energy spread after BEMS upgrade.}
\begin{center}
\begin{tabular}{|c|c|c|} \hline \hline
     & Positron & Electron \\ \hline
  Energy (MeV) & 1117.134 $\pm$ 0.071 & 1116.763 $\pm$ 0.039\\ \hline
 Energy spread (keV) & 726.9 $\pm$ 102.6 & 741.1 $\pm$ 50.6\\ \hline
\end{tabular}
\end{center}
\begin{multicols}{2}

Two typical measured results with all upgraded components are shown in Figure~\ref{result}, where the Compton edges of electron and positron are measured April 2015 during the data taking of Y(2235) sample. The results are listed in Table ~\ref{info}, it is clear that twelve minutes data collection, the Compton edge is very sharp, the measurement precision of beam energy is about 6.5$\times$ 10$^{-5}$, the energy spread is better than 15\%.

\begin{center}
\begin{minipage}{4cm}
\includegraphics[width=4.cm]{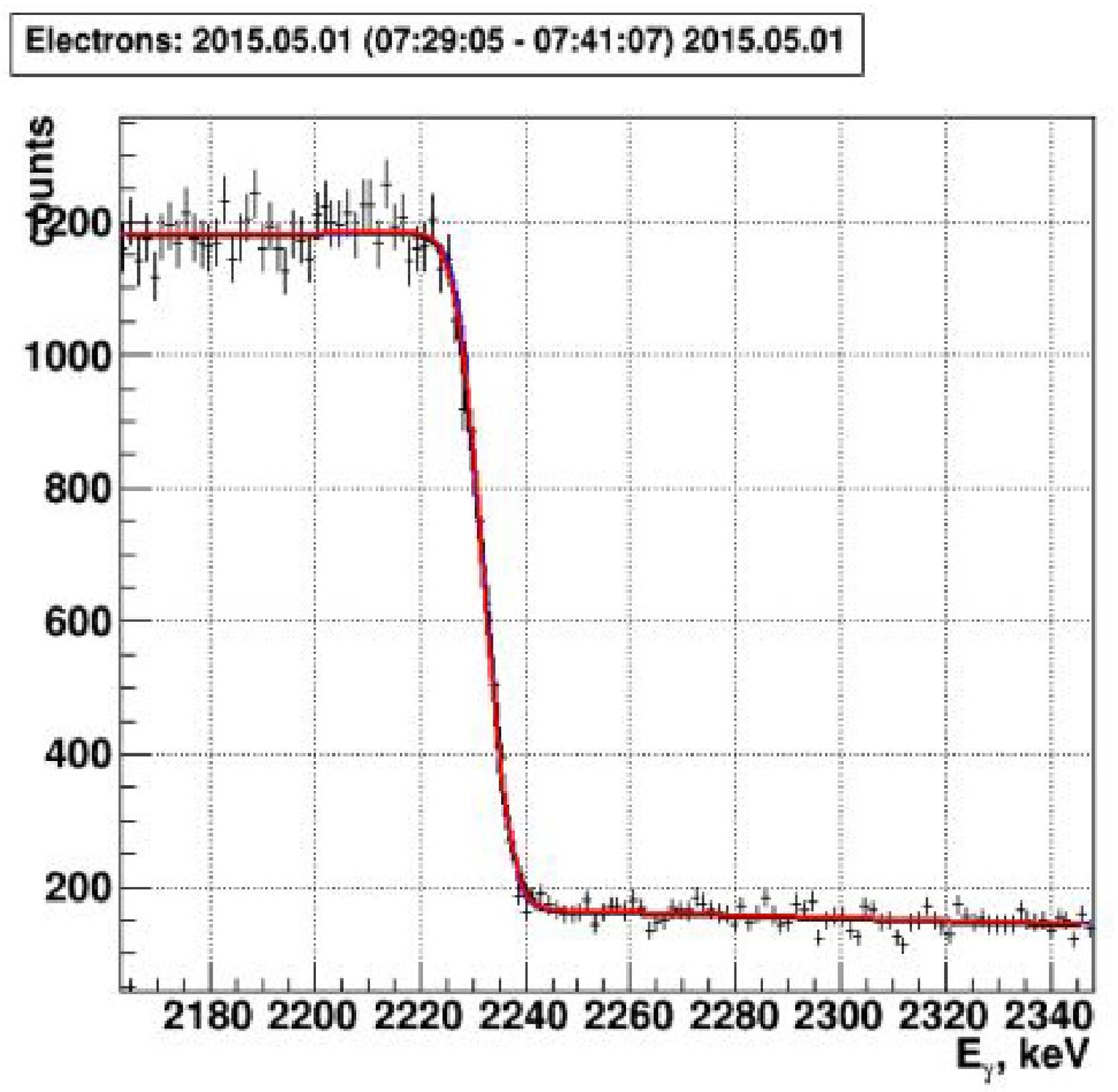}
\centerline{(a)}
\end{minipage}
\begin{minipage}{4cm}
\includegraphics[width=4.cm]{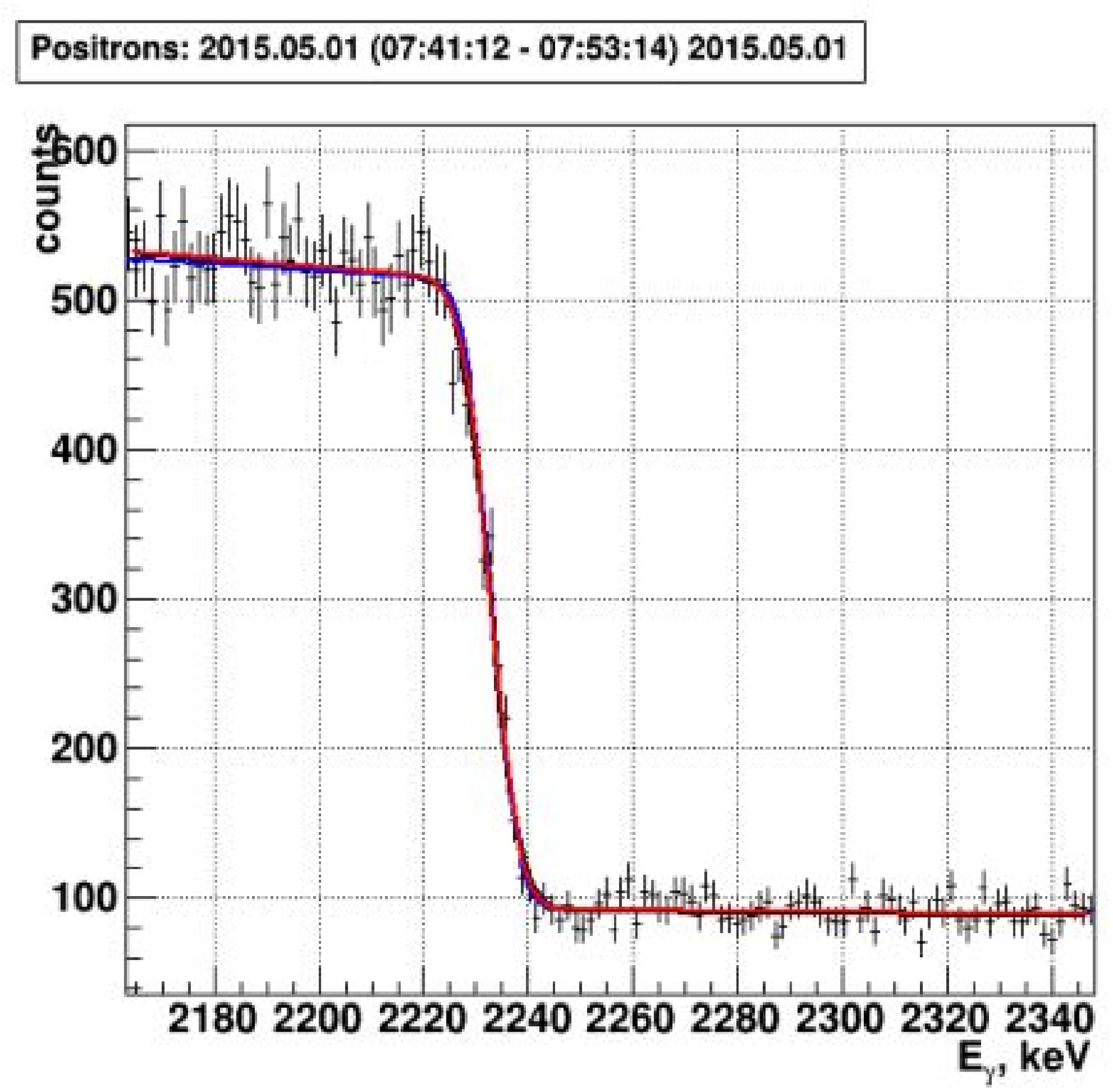}
\centerline{(b)}
\end{minipage}
\figcaption{\label{result}
(a): Compton edge for electron case. (b): Compton edge for positron case.}
\end{center}

\vspace{-1mm}
\centerline{\rule{80mm}{0.1pt}}
\vspace{2mm}


\end{multicols}

\clearpage
\end{CJK*}
\end{document}